\documentclass[%
reprint,
superscriptaddress,
nofootinbib,
amsmath,amssymb,
aps,
noeprint,
]{revtex4-2}
\usepackage[normalem]{ulem}
\usepackage[utf8]{inputenc}
\usepackage[T1]{fontenc}
\usepackage{bm}
\usepackage[separate-uncertainty=true]{siunitx}
\usepackage{booktabs}
\usepackage{textgreek}
\usepackage{upgreek}
\usepackage{subfigure}
\usepackage{mhchem}
\usepackage{ulem}
\usepackage{autobreak}

\usepackage{lineno}
\usepackage{todonotes}  
\usepackage{amsmath,amssymb}
\usepackage{pifont}

\newcommand{\phaseSpace}[2]{{#1} \sqrt{{#1}^2 - #2^2}}

\DeclareMathOperator{\heavyside}{\Theta}
\newcommand{\neutrinoEnergy}{\varepsilon_f}

\newcommand{\reducedEndpoint}{{E_0}}

\newcommand{\cabibbo}{{\Theta_\mathrm{C}}}
\newcommand{\couplingGen}[2]{{#1}_\mathrm{#2}}

\newcommand{\Coupling}[1]{\couplingGen{G}{#1}}
\newcommand{\couplingSqGen}[2]{\couplingGen{#1}{#2}^2}

\newcommand{\CouplingSq}[1]{\couplingSqGen{G}{#1}}

\newcommand{\mnu}{\mbox{$m_{\nu}$}}
\newcommand{\mnutwo}{\mbox{$m_{\nu}^2$}}

\newcommand{\mnui}{\mbox{$m_\mathrm{i}$}}

\begin{document}
\title{First direct neutrino-mass measurement with sub-eV sensitivity}

\newcommand{\berlin}{Institut f\"{u}r Physik, Humboldt-Universit\"{a}t zu Berlin, Newtonstr. 15, 12489 Berlin, Germany}
\newcommand{\bonn}{Helmholtz-Institut f\"{u}r Strahlen- und Kernphysik, Rheinische Friedrich-Wilhelms-Universit\"{a}t Bonn, Nussallee 14-16, 53115 Bonn, Germany}
\newcommand{\cmu}{Department of Physics, Carnegie Mellon University, Pittsburgh, PA 15213, USA}
\newcommand{\cwru}{Department of Physics, Case Western Reserve University, Cleveland, OH 44106, USA}
\newcommand{\etp}{Institute of Experimental Particle Physics~(ETP), Karlsruhe Institute of Technology~(KIT), Wolfgang-Gaede-Str. 1, 76131 Karlsruhe, Germany}
\newcommand{\fulda}{University of Applied Sciences~(HFD)~Fulda, Leipziger Str.~123, 36037 Fulda, Germany}
%
%
%
\newcommand{\iap}{Institute for Astroparticle Physics~(IAP), Karlsruhe Institute of Technology~(KIT), Hermann-von-Helmholtz-Platz 1, 76344 Eggenstein-Leopoldshafen, Germany}
\newcommand{\ipe}{Institute for Data Processing and Electronics~(IPE), Karlsruhe Institute of Technology~(KIT), Hermann-von-Helmholtz-Platz 1, 76344 Eggenstein-Leopoldshafen, Germany}
\newcommand{\itep}{Institute for Technical Physics~(ITEP), Karlsruhe Institute of Technology~(KIT), Hermann-von-Helmholtz-Platz 1, 76344 Eggenstein-Leopoldshafen, Germany}
\newcommand{\tlk}{Tritium Laboratory Karlsruhe~(TLK), Karlsruhe Institute of Technology~(KIT), Hermann-von-Helmholtz-Platz 1, 76344 Eggenstein-Leopoldshafen, Germany}
\newcommand{\ppq}{Project, Process, and Quality Management~(PPQ), Karlsruhe Institute of Technology~(KIT), Hermann-von-Helmholtz-Platz 1, 76344 Eggenstein-Leopoldshafen, Germany    }
%
%
\newcommand{\inr}{Institute for Nuclear Research of Russian Academy of Sciences, 60th October Anniversary Prospect 7a, 117312 Moscow, Russia}
\newcommand{\lbnl}{Institute for Nuclear and Particle Astrophysics and Nuclear Science Division, Lawrence Berkeley National Laboratory, Berkeley, CA 94720, USA}
\newcommand{\madrid}{Departamento de Qu\'{i}mica F\'{i}sica Aplicada, Universidad Autonoma de Madrid, Campus de Cantoblanco, 28049 Madrid, Spain}
\newcommand{\mainz}{Institut f\"{u}r Physik, Johannes-Gutenberg-Universit\"{a}t Mainz, 55099 Mainz, Germany}
\newcommand{\mpp}{Max-Planck-Institut f\"{u}r Physik, F\"{o}hringer Ring 6, 80805 M\"{u}nchen, Germany}
\newcommand{\massit}{Laboratory for Nuclear Science, Massachusetts Institute of Technology, 77 Massachusetts Ave, Cambridge, MA 02139, USA}
\newcommand{\mpik}{Max-Planck-Institut f\"{u}r Kernphysik, Saupfercheckweg 1, 69117 Heidelberg, Germany}
\newcommand{\muenster}{Institut f\"{u}r Kernphysik, Westf\"alische Wilhelms-Universit\"{a}t M\"{u}nster, Wilhelm-Klemm-Str. 9, 48149 M\"{u}nster, Germany}
\newcommand{\npi}{Nuclear Physics Institute of the CAS, v. v. i., CZ-250 68 \v{R}e\v{z}, Czech Republic}
\newcommand{\unc}{Department of Physics and Astronomy, University of North Carolina, Chapel Hill, NC 27599, USA}
\newcommand{\washington}{Center for Experimental Nuclear Physics and Astrophysics, and Dept.~of Physics, University of Washington, Seattle, WA 98195, USA}
\newcommand{\wuppertal}{Department of Physics, Faculty of Mathematics and Natural Sciences, University of Wuppertal, Gau{\ss}str. 20, 42119 Wuppertal, Germany}
\newcommand{\saclay}{IRFU (DPhP \& APC), CEA, Universit\'{e} Paris-Saclay, 91191 Gif-sur-Yvette, France }
\newcommand{\tum}{Technische Universit\"{a}t M\"{u}nchen, James-Franck-Str. 1, 85748 Garching, Germany}
\newcommand{\tunl}{Triangle Universities Nuclear Laboratory, Durham, NC 27708, USA}
%
%
\newcommand{\ornl}{Also affiliated with Oak Ridge National Laboratory, Oak Ridge, TN 37831, USA}
%
%
%



\author{M.~Aker}\affiliation{\tlk}
\author{A.~Beglarian}\affiliation{\ipe}
\author{J.~Behrens}\affiliation{\etp}\affiliation{\iap}
\author{A.~Berlev}\affiliation{\inr}
\author{U.~Besserer}\affiliation{\tlk}
\author{B.~Bieringer}\affiliation{\muenster}
\author{F.~Block}\affiliation{\etp}
\author{B.~Bornschein}\affiliation{\tlk}
\author{L.~Bornschein}\affiliation{\iap}
\author{M.~B\"{o}ttcher}\affiliation{\muenster}
% ...

\begin{abstract}
We report the results of the second measurement campaign of the Karlsruhe Tritium Neutrino (KATRIN) experiment. KATRIN probes the effective electron anti-neutrino mass, \mnu, via a high-precision measurement of the tritium $\upbeta$-decay spectrum close to its endpoint at \SI{18.6}{\kilo\electronvolt}. In the second physics run presented here, the source activity was increased by a factor of 3.8 and the background was reduced by $25\,\%$ with respect to the first campaign. A sensitivity on \mnu\  of \SI{0.7}{\electronvolt\per c \squared} at \SI{90}{\percent} confidence level (CL) was reached. This is the first sub-eV sensitivity from a direct neutrino-mass experiment. The best fit to the spectral data yields $\mnutwo =\SI{0.26\pm0.34}{\electronvolt\squared\per c \tothe{4}}$, resulting in an upper limit of $m_{\nu}<\SI{0.9}{\electronvolt\per c \squared}$ (\SI{90}{\percent}  CL). By combining this result with the first neutrino mass campaign, we find an upper limit of $\mnu<\SI{0.8}{\electronvolt\per c \squared}$ (\SI{90}{\percent}  CL).
\end{abstract}

\flushbottom
\maketitle

\thispagestyle{empty}

\section{Introduction}
The discovery of neutrino flavour oscillations~\cite{Fukuda:1998mi,Ahmad:2002jz} proves that neutrinos must have a mass, unlike originally assumed in the Standard Model (SM) of Particle Physics. Neutrino oscillation experiments have shown that the weakly interacting neutrino flavour eigenstates, $\nu_{f}$ with $f \in \{\text{e},\upmu,\uptau\}$, are admixtures of the three neutrino mass eigenstates $\nu_i$ with mass eigenvalues $m_i$. While neutrino-oscillation experiments can probe the differences of squared neutrino mass eigenvalues $\Delta m_{ij}^2 $, the absolute neutrino-mass scale remains one of the most pressing open questions in the fields of nuclear, particle, and astroparticle physics today. In this paper we report a measurement of the effective electron anti-neutrino mass defined as $\mnutwo=\sum_i |U_{\mathrm{e}i}|^2 m_i^2$ where $U_{\mathrm{e}i}$ are elements of the Pontecorvo-Maki-Nakagawa-Sakata (PMNS) matrix, which describes the mixing of the neutrino states.

The neutrino masses are at least five orders of magnitude smaller than the mass of any other fermion of the SM, which may point to a different underlying mass-creation mechanism~\cite{Petcov:2013}. The determination of the neutrino mass would thus shed light on the fundamental open question of the origin of particle masses. Despite the smallness of their masses, neutrinos play a crucial role in the evolution of large-scale structures of our cosmos due to their high abundance in the universe \cite{10.1111/j.1365-2966.2010.17546.x,LESGOURGUES2006307}. A direct measurement of the neutrino mass could hence provide a key input to cosmological structure-formation models. In this respect, cosmological observations themselves provide stringent limits on the sum of neutrino masses\footnote{In this paper we use the convention $c=1$.} of $\sum m_i<\SI{0.12}{\electronvolt}$ (\SI{95}{\percent}  CL)~\cite{2021PhRvD.103h3533A,Aghanim:2018eyx}. However, these limits strongly rely on the underlying cosmological assumptions~\cite{PhysRevD.92.121302,Boyle:2017lzt}. An independent measurement of {the neutrino mass} could help to break parameter degeneracies of the cosmological models. A powerful way to probe {this neutrino property} in the laboratory is via a search for neutrinoless double $\upbeta$-decay, providing limits at the level of $m_{\upbeta\upbeta}<\SI{0.08}-\SI{0.18}{\electronvolt}$ (\SI{90}{\percent} CL)~\cite{Agostini:2020xta,PhysRevLett.117.082503}, depending on the nuclear matrix element calculation. In contrast to \mnu, the effective mass in double-beta decay is given by $m_{\upbeta\upbeta}=\left|\sum_i U_{\mathrm{e}i}^2 m_i\right|$. The limit is only valid under the assumption that neutrinos are their own anti-particle (Majorana particle) and that light neutrinos mediate the decay.   

The most direct way to assess the neutrino mass is via the kinematics of a single $\upbeta$-decay. This method is independent of any cosmological model and of the mass nature of the neutrino, i.e. it may be a lepton of Majorana or Dirac type. The neutrino masses \mnui ~lead to a reduction of the maximal observed energy of the decay and a small spectral shape distortion close to the kinematic endpoint of the $\upbeta$-spectrum. In the quasi-degenerate mass regime, where $m_i>\SI{0.2}{\electronvolt}$, the mass splittings are negligible with respect to the masses, $m_i$, and the observable can be approximated as $\mnutwo=\sum_i |U_{\mathrm{e}i}|^2 m_i^2$.


The Karlsruhe Tritium Neutrino (KATRIN) experiment~\cite{KDR2004,aker2021design} exploits the single $\upbeta$-decay of molecular tritium
\begin{align}
 \ce{T2} \to \ce{^3HeT+} + \mathrm{e}^- + \bar{\nu}_e   
\end{align}
and currently provides the best neutrino mass sensitivity in the field of direct neutrino-mass measurements with its first published limit of $\mnu < \SI{1.1}{\electronvolt}$ (\SI{90}{\percent} CL)~\cite{Aker:2019uuj,aker2021analysis}. KATRIN is designed to determine the neutrino mass with a sensitivity of close to $\SI{0.2}{\electronvolt}$ (\SI{90}{\percent} CL) in a total measurement time of 1000 days~\cite{KDR2004}. In this work, we present the second neutrino-mass result of KATRIN, reaching an unprecedented sub-eV sensitivity and limit in $\mnu$ from a direct measurement. 

\section{Experimental setup}
{
The design requirements to detect the small signature of a neutrino mass in the last few eV of a $\upbeta$-decay spectrum are: a high tritium activity ($\SI{1e11}{\becquerel}$), a low background rate ($\lesssim$ 0.1 counts per second (cps)), an eV-scale energy resolution, and an accurate (\SI{0.1}{\percent}-level) theoretical prediction of the integral spectrum.}

\begin{figure*}
    \centering
    \includegraphics[width=0.85\textwidth]{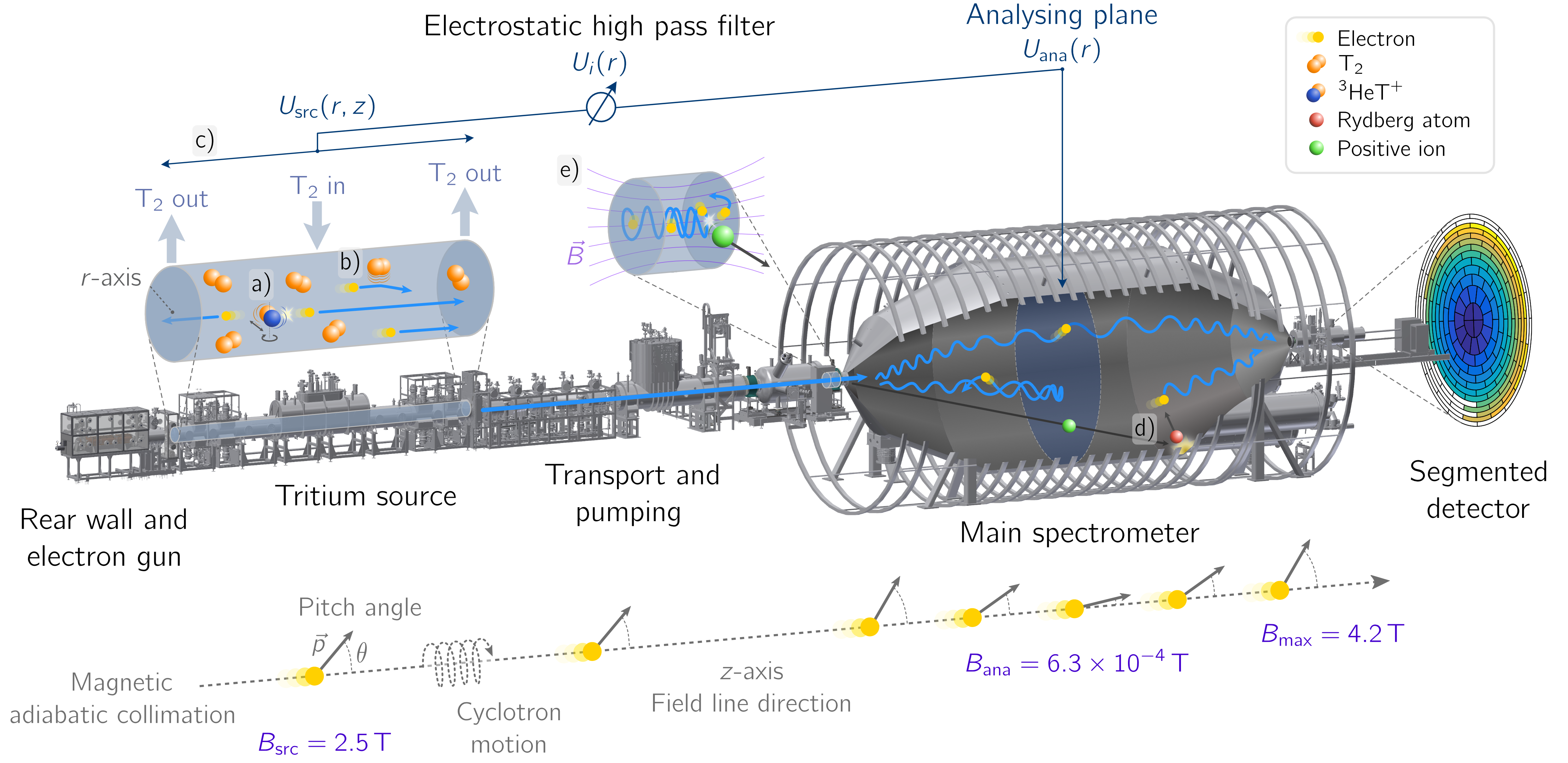}
    \caption{
    \small{Working principle of the KATRIN experiment as explained in the main text. The view into the tritium source depicts three systematic effects: a) molecular excitations during $\upbeta$-decay, b) scattering of electrons off the gas molecules, and c) spatial distribution of the electric potential in the source $U_{\mathrm{src}}(r,z)$. The view into the spectrometer illustrates the main background processes. Generally, unstable neutral particles (d) can enter the large spectrometer volume and decay. The resulting electrons are accelerated by $qU_{\mathrm{ana}}$ towards the detector, making them indistinguishable from signal electrons~\cite{Fraenkle:2017zpo, Aker:2019uuj}. The unstable neutral particles are either $^{219}$Rn and $^{220}$Rn emanated from the getter pumps~\cite{Frankle:2011xy,Mertens:2012vs} or highly excited atoms sputtered off the inner spectrometer surfaces~\cite{TrostThesis,Schaller:2020kem,fraenkle2020katrin}. The sputtering mechanisms originate from both $\alpha$-decays of $^{210}$Po in the structural material and a Penning trap (e), which gives rise to accelerated positive ions from the volume between the two spectrometers~\cite{KATRIN-Penning:2020}.} Note that the white pixels in the detector were excluded from analysis as they do not fulfil the `golden pixel' quality criteria. {The magnetic adiabatic collimation is only sketched for the case without electric field for illustration purposes.}}
    \label{fig:experiment}
\end{figure*}

The KATRIN experiment tackles these challenges by combining a high-activity molecular tritium source with a high-resolution spectrometer of the \textit{Magnetic Adiabatic Collimation and Electrostatic} (MAC-E)-Filter type \cite{aker2021design}. The experiment is hosted by the Tritium Laboratory Karlsruhe (TLK) allowing the safe supply of tritium at the 10-gramme scale with continuous tritium reprocessing~\cite{Priester:2015bfa, Priester:FST2020}. Fig.\ \ref{fig:experiment} shows the 70-m long KATRIN beamline.
The isotope tritium has a short half life of 12.3\ years and a low endpoint of $\SI{18.6}{\kilo\electronvolt}$, both favourable properties to achieve high count rates near the endpoint. Moreover, the theoretical calculation of the super-allowed decay of tritium is well understood~\cite{Wilkinson:1991kog,Simkovic:2007yi,Masood:2007rc}. The strong gaseous tritium source (nominal activity $\SI{1e11}{\becquerel}$) is established by the throughput of \SI{40}{\gram}/day of molecular tritium through the 10-m long source tube, which is cooled to $\SI{30}{\kelvin}$ to reduce thermal motion of the tritium molecules.

A system of 24 superconducting magnets~\cite{Arenz:2018jpa} guides the electrons out of the source towards the spectrometer section and to the detector. Between the source and spectrometer, differential~\cite{Marsteller2021} and cryogenic~\cite{Gil2010} pumping systems reduce the flow of tritium by 14 orders of magnitude.

High-precision electron spectroscopy is achieved with the spectrometer of MAC-E-filter type~\cite{Lobashev:1985mu,Picard1992}. The spectrometer acts as a sharp electro-static high-pass filter, transmitting only electrons (of charge $q=-e$) above an adjustable energy threshold $qU$, {where $U$ is the retarding potential applied at the spectrometer.} A reduction of the magnetic field strength by about four orders of magnitudes from the entrance (exit) of the spectrometer $B_\mathrm{source} = \SI{2.5}{\tesla}$ ($B_\mathrm{max} = \SI{4.2}{\tesla}$) to its center, the analysing plane ($B_\mathrm{ana} = \SI{6.3e-4}{\tesla}$), collimates the electron momenta. This configuration creates a narrow filter width of $\Delta E=\SI{18.6}{\kilo\electronvolt}\cdot (B_\mathrm{ana}/B_\mathrm{max})=\SI{2.8}{\electronvolt}$ and allows for a large angular acceptance, with a maximum pitch-angle~\footnote{The pitch angle refers to the angle between the electron's momentum and the direction of the magnetic field at the position of the electron.} for the $\upbeta$-decay electrons of $\theta_{\mathrm{max}} = \arcsin{\sqrt{(B_\mathrm{source}/B_\mathrm{max})}} = 50.4^{\circ}$ in the source. A 12-m diameter magnetic coil system surrounding the spectrometer fine-shapes the magnetic field and compensates the Earth's magnetic field~\cite{Glueck2013, Erhard:2017htg}. 

The transmitted electrons are detected by a 148-pixel silicon PIN-diode focal-plane detector installed at the exit of the spectrometer~\cite{Amsbaugh:2014uca}. By measuring the count rate of transmitted electrons for a set of $qU$ values, the integral $\upbeta$-decay spectrum is obtained. The main spectrometer is preceded by a smaller pre-spectrometer, which operates on the same principle and transmits only electrons above \SI{10}{\kilo\electronvolt}, to reduce the flux of electrons into the main spectrometer. 
The up-stream end of the beamline is terminated with a gold-plated rear wall, which absorbs the non-transmitted $\upbeta$-electrons and defines the reference electric potential of the source. The rear wall is biased to a voltage ($\mathcal{O}(\SI{100}{\milli\volt})$) to minimise the difference of its surface potential to that of the beam tube, which minimises the inhomogeneity of the source electric potential.

The rear section is equipped with an angular- and energy-selective electron gun~\cite{Behrens:2017cmd}, which is used to precisely determine the scattering probability of electrons with the source gas, governed by the product of column density (number of molecules per cm$^2$ along the length of the source) and scattering cross section. 
Furthermore, we use the electron gun to measure the distribution of energy losses for \SI{18.6}{\kilo\electronvolt} electrons scattering off the molecular tritium gas, providing the most precise energy loss function for this process to date ~\cite{aker2021precision}.
Another key calibration source is gaseous krypton, which can be co-circulated with the tritium gas~\cite{Sentkerestiova-GKrS:2018}. Mono-energetic conversion electrons from the decay of the meta-stable state ${}^{83\mathrm{m}}$Kr are used to determine spatial and temporal variations of the electric potential in the tritium source. The variations are caused by a weak cold-magnetised plasma, which arises from the high magnetic field ($\SI{2.5}{\tesla}$) and a large number of ions and low-energy electrons ($\approx \SI{1e12}{\per\cubic\metre}$) in the tritium source. The methods of calibration are described in more detail in Sec.\ \ref{sec:methods}.

The beamline is equipped with multiple monitoring devices. A laser Raman system continuously monitors the gas composition, providing a measurement at the $\SI{0.1}{\percent}$-precision level each minute. A silicon drift detector system, installed in the transport section, as well as a beta-induced X-ray system at the rear section~\cite{Roellig2015}, continuously monitor the tritium activity, obtaining a result at the $\SI{0.1}{\percent}$-precision level each minute. The high voltage of the main spectrometer is continuously measured at the ppm level with a high-precision voltage divider system~\cite{Arenz:2018ymp,Thummler:2009rz} and an additional monitoring spectrometer~\cite{Erhard2014}. The magnetic fields are determined with a high-precision magnetic field sensor system~\cite{Letnev:2018fkq}. 

\begin{figure}
    \centering
    \includegraphics[width=\linewidth]{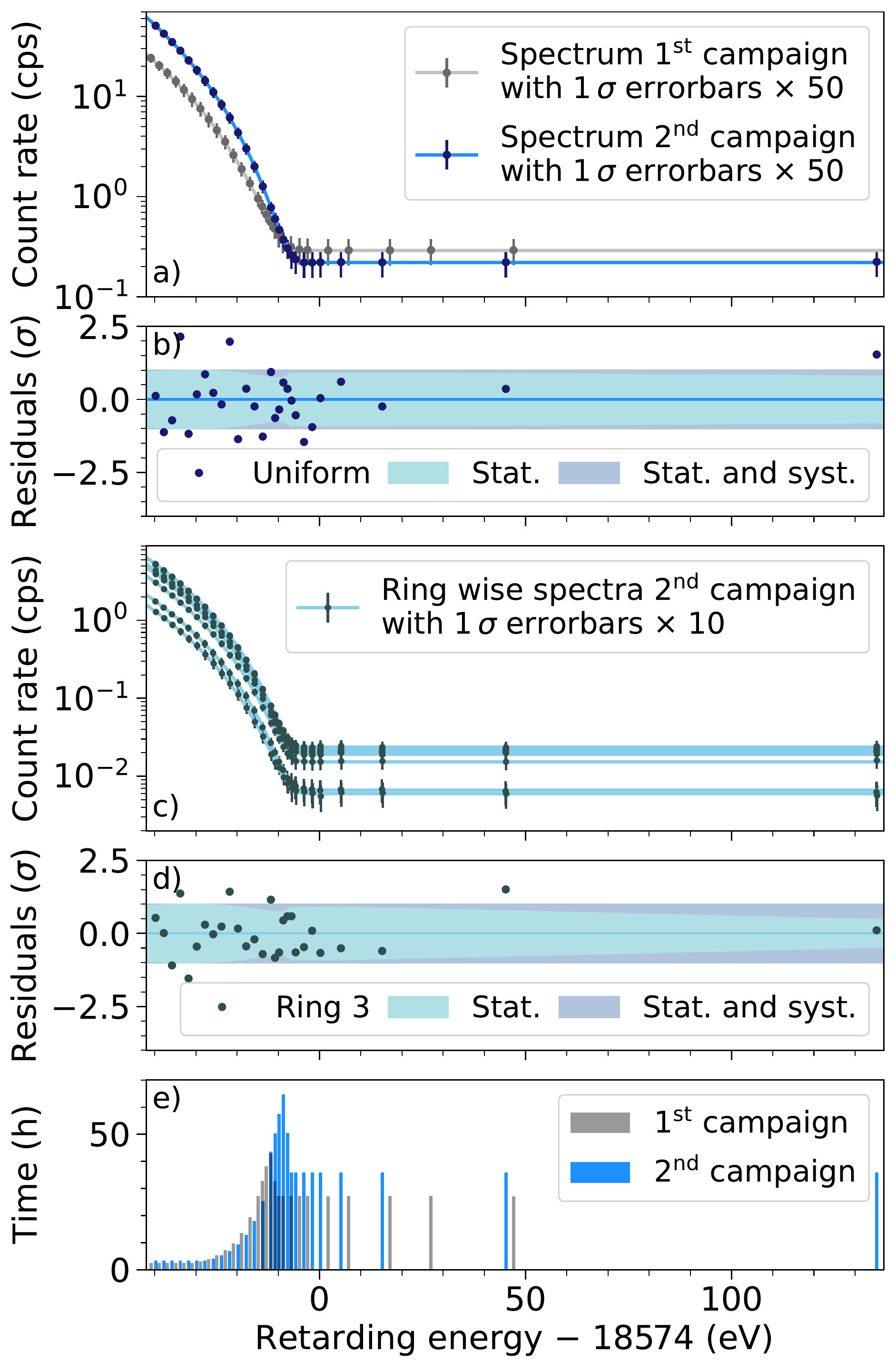}
    \caption{a): Individual fits to both KATRIN measurement campaigns (KNM1 and KNM2). Here the 12 detector rings are combined to a single detector area (uniform fit). The graph illustrates the improvements in lower background rate, higher signal strength and overall better statistics (as indicated by smaller error bars). b) Normalised residuals for the uniform fit of the KNM2 data set. c) Simultaneous fit of the rates measured with 12 detector rings {(note that most of the ring overlap)} at 28 retarding energies in the KNM2 measurement campaign (see main text).
    d) Normalised residuals for the fit to the data of the (exemplary) third detector ring of the KNM2 data set. e) Measurement time distribution for KNM1 and KNM2.}
    \label{fig:fit}
\end{figure}

After the successful commissioning of the complete KATRIN beamline in summer 2017~\cite{Arenz:2018kma}, first tritium operation was demonstrated with a small tritium activity of (\SI{5e8}{\becquerel}) in mid-2018~\cite{Aker:2019qfn}. During the first neutrino mass campaign (KNM1) in 2019~\cite{Aker:2019uuj}, the source was operated in a `burn-in' configuration at a reduced activity of $\SI{2.5e10}{\becquerel}$, which is required when structural materials are exposed to high amounts of tritium for the first time. Major technical achievements of the second measurement campaign (KNM2) are the operation of the tritium source at its nominal activity of $\SI{9.5e10}{\becquerel}$ and improved vacuum conditions in the spectrometer~\cite{Arenz:2016mrh} that led to a reduction of the background by $\SI{25}{\percent}$ to $\SI{220}{mcps}$. We thus increased the $\upbeta$-electron-to-background ratio by a factor of 2.7 with respect to the first campaign. In the last $\SI{40}{\electronvolt}$ of the integral spectrum, we collected a total number of $\SI{3.7e6}{}$ $\upbeta$-electrons. Fig.\ \ref{fig:fit} a) compares the spectra of both neutrino mass campaigns. A direct comparison of the experimental parameters is given in Tab.\ \ref{tab:comparison}.




\begin{table*}[]
    \centering
    \caption{Comparison of key numbers for KATRIN Neutrino Mass (KNM) campaigns. KNM1 refers to the first KATRIN results~\cite{Aker:2019uuj} and KNM2 to this work. The total number of $\upbeta$-electrons is counted in the last $\SI{40}{\electronvolt}$ interval of the integral spectrum, which is used for the spectral fit. The $\upbeta$-electron-to-background ratio is given by the ratio of this number and the background counts in the same energy range, i.~e. $E_0-\SI{40}{\electronvolt}$ to $E_0$.}
    \begin{tabular}{llll}
    \toprule
          & KNM1 & KNM2 \\
        \midrule
        Number of scans                 & 274       & 361 \\
        Total scan time                 & \SI{521.7}{\hour}   & \SI{743.7}{\hour} \\ 
        Background rate                 & \SI{290}{mcps} & \SI{220}{mcps}\\
        T$_{2}$ column density          & \SI{1.11e17}{\per\centi\metre\squared}&  \SI{4.23e17}{\per\centi\metre\squared}\\
        Source activity                 & \SI{2.5e10}{\becquerel} & \SI{9.5e10}{\becquerel} \\
        Total number of $\upbeta$-electrons & $\SI{1.48e6}{}$     & $\SI{3.68e6}{}$\\
        $\upbeta$-electron-to-background ratio & 3.7 & 9.9\\
         \bottomrule
    \end{tabular}
    \label{tab:comparison}
\end{table*}

\section{Measurement of the tritium beta spectrum}
The integral $\upbeta$-spectrum is obtained by repeatedly measuring the count rate $R_{\mathrm{data}}(qU_i)$ for a set of 39 non-equidistant high voltage (HV) settings $U_i$, creating a retarding energy $qU_i$ for the electrons of charge $q$. The retarding energy is adjusted in a range of $qU_i\in \left(E_{0} - \SI{300}{\electronvolt}, E_{0} + \SI{135}{\electronvolt}\right)$, where $E_{0}=\SI{18574}{\electronvolt}$ is the approximate spectral endpoint. Note that for the spectral fit, only 28 of those points in {the} range of $qU_i\in[E_{0} - \SI{40}{\electronvolt}, E_{0} + \SI{135}{\electronvolt}]$ are used. Data points further below the endpoint are used to monitor the activity stability, complementing the other monitoring devices mentioned above. The time spent at each HV set-point (called \textit{scan step}) varies between $\SI{17}{\second}$ and $\SI{576}{\second}$ and is chosen to optimise the neutrino mass sensitivity. The total time to measure the rate at all 39 retarding energies (a so-called \textit{scan}) is about \SI{2}{\hour}. As can be seen in Fig.\ \ref{fig:fit} e), the measurement time distribution (MTD) compensates for the steeply falling count rate towards the endpoint and peaks at approximately $\SI{10}{\electronvolt}$ below the endpoint, where a neutrino-mass signal would be maximal. Based on pre-defined experimental {conditions} and data quality criteria, 361 golden scans were chosen for the presented analysis. Since the HV values can be set with a reproducibility at the sub-ppm-level, these scans are later combined to a single spectrum by adding the counts at a given set point.

The focal-plane detector, shown as magnified inset in Fig.\ \ref{fig:experiment}, is segmented into 148 individual pixels to account for spatial variations of the electromagnetic fields inside the source and spectrometer, and of the background. Removing malfunctioning pixels and those shadowed by hardware upstream, 117 pixels have been selected. For the presented analysis these golden pixels are grouped into 12 concentric rings, resulting in 336 data points $R_\mathrm{data}(qU_i,r_j)$ for $i \in \{1, \ldots, n_{\text{qU}}=28\}$ and $j \in \{1, \ldots, n_{\text{rings}}=12\}$ per scan. Fig.\ \ref{fig:fit} c) displays the spectra for each of the 12 detector rings, where all scans have been combined.

\section{Data analysis}
In order to infer the neutrino mass, we fit the spectral data with a spectrum prediction, given by an analytical description of the $\upbeta$-decay spectrum and the experimental response function, described in the following sections.

\subsection{Spectrum calculation}
The prediction of the detection rate $R(qU_i,r_j)$ is given by a convolution of the differential $\upbeta$-decay spectrum $R_\upbeta(E)$ with the experimental response function $f(E, qU_i, r_j)$, and a background rate $R_{\mathrm{bg}}(qU_i,r_j)$:
\begin{small}
\begin{align}
\label{Ntheo}
R(qU_i, r_j) & = A_{\mathrm{s}} N_{\mathrm{T}} \int_{qU_i}^{E_0}R_\upbeta(E) f(E, qU_i, r_j) dE + R_{\mathrm{bg}}(qU_i, r_j). 
\end{align}
\end{small}
Here $N_{\mathrm{T}}$ is the signal normalisation calculated from the number of tritium atoms in the source, the maximum acceptance angle, and the detection efficiency. $A_{\mathrm{s}}\approx 1$ is an additional normalisation factor, which is used as a free parameter in the fit. The differential $\upbeta$-decay spectrum $R_\upbeta(E)$ is given by Fermi's theory. In the analysis we include radiative corrections, and the molecular final-state distribution (FSD)~\cite{Saenz2000,Doss2006} in the differential $\upbeta$-decay spectrum. The Doppler broadening due to the finite thermal motion the tritium molecules in the source, as well as energy broadenings due to spatial and temporal variations of the spectrometer and source electric potential, are emulated by Gaussian broadenings of the FSD. 

The response function $f(E, qU_i, r_j)$ includes the energy losses due to scattering and synchrotron radiation in the high $B$-field regions, as well as the transmission of electrons through the main spectrometer. Compared to previous KATRIN analyses, we now use a modified transmission function to account for the non-isotropy of the $\upbeta$-electrons leaving the source\footnote{The angular distribution of electrons leaving the source is slightly non-isotropic, due to the pitch-angle dependence of the scattering probabilities, which arises from the pitch-angle dependence of the effective path lengths through the source.}. A detailed description of the spectrum calculation can be found in Ref.~\cite{Kleesiek:2018mel} and in section~\ref{sec:methods}.


\subsection{Unbiased parameter inference}


We infer \mnutwo\ by fitting the experimental spectrum $R_\mathrm{data}(qU_i, r_j)$ with the prediction $R(qU_i, r_j)$ by minimising the standard $\chi^2=\vec{R}_\mathrm{data} C^{-1} \vec{R}$ function, where $C$ contains the variance and co-variance of the data points. In addition to the neutrino mass squared, $\mnutwo$, the parameters $A_{\mathrm{s}}(r_j)$, $R_{\mathrm{bg}}(r_j)$, and the effective endpoint $E_0(r_j)$ are treated as independent parameters for the 12 detector rings, leading to a total number of free parameters of $1+3\times12=37$ in the fit. 
The introduction of ring-dependent parameters was chosen to allow for possible unaccounted radial effects. In particular, the effective endpoint $E_0(r_j)$ could account for a possible radial dependence of electric potential in the source. However, the final fit revealed a negligible (<\SI{100}{\milli\electronvolt}) radial variation of the endpoint. 
Another advantage of ring-dependent parameters is to avoid the averaging of the transmission function over all rings, which would introduce an energy broadening, and hence reduce the resolution.


The following analysis procedure was implemented to minimise the potential for human-induced biases. The full analysis is first performed on a Monte-Carlo copy of each scan, simulated based on the true experimental parameters and with the neutrino mass set to zero. After all systematic inputs (e.g. the magnetic field values and uncertainties) are quantified, the fit is performed on the experimental data set, but with a randomly broadened molecular final-state distribution, which imposes an unknown bias on the observable $\mnutwo$. Only after three independent analysis teams, using different software and strategies, obtained consistent results would the analysis of the true data with unmodified final-state distribution be performed.

\subsection{Systematic effects}
The analytical description $R(qU_i, r_j)$ of the integral $\upbeta$-spectrum contains various experimental and theoretical parameters, such as the magnetic fields, the column density, and the probability for given molecular excitations, which are known with a certain accuracy. Different techniques (based on covariance matrices, Monte-Carlo propagation, nuisance parameters, and Bayesian priors) are applied to propagate these systematic uncertainties in the final result. A detailed description of these methods can be found in Sec.\ \ref{sec:methods}.

The neutrino mass result presented here is dominated by the statistical uncertainty. The largest systematic uncertainties are related to background properties and the source electric potential. First, radon decays in the main spectrometer lead to a non-Poissonian background rate over-dispersion~\cite{Mertens:2012vs} of about \SI{11}{\percent}, effectively increasing the statistical uncertainty. Secondly, background generation mechanisms may show a retarding-potential dependence of the background, parametrize by a slope ($m^{qU}_{\mathrm{bg}} = \SI{0 \pm 4.74}{mcps\per\kilo\electronvolt}$.) . Thirdly, a removal of stored electrons from a known Penning trap between the spectrometers~\cite{KATRIN-Penning:2020} after each scan step can lead to a slowly increasing background rate ($m^{\mathrm{t}_{\mathrm{scan}}}_{\mathrm{bg}} = \SI{3 \pm 3}{\micro cps\per\second}$) and thus to a scan-step-duration-dependent background contribution. Finally, spatial and temporal variations of the source electric potential modify the spectral shape and therefore lead to a relevant systematic uncertainty for the neutrino mass measurement. The impact of all systematic effects on the neutrino mass is listed in Tab.\ \ref{tab:systematics_breakdown-final} and described in detail in Sec.\ \ref{sec:methods}. 


\begin{figure}[]
    \includegraphics[width=\linewidth]{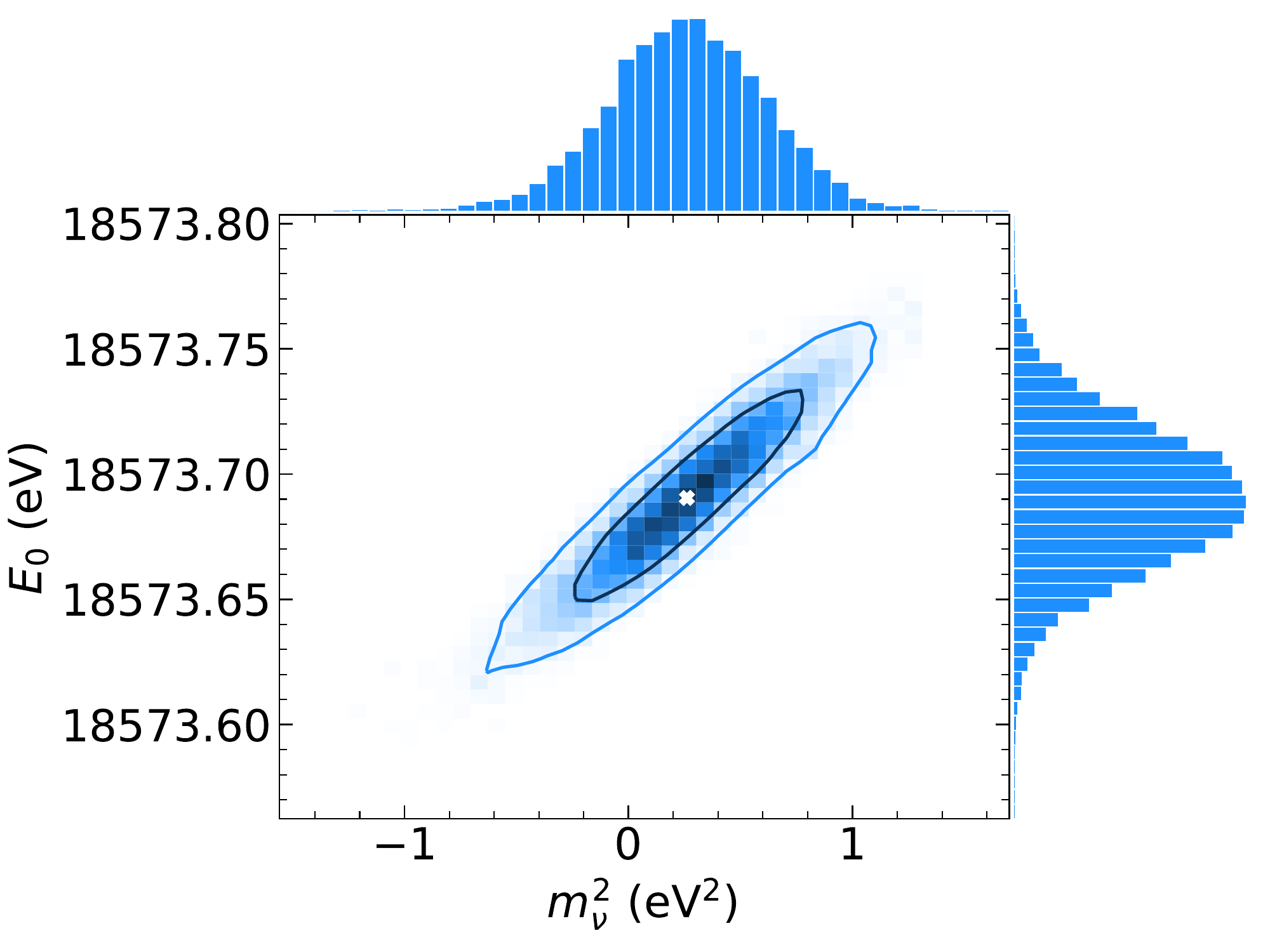}
    \caption{Distribution of fitted $\mnutwo$~and $E_0$ values, obtained by the Monte-Carlo propagation technique (see section~\ref{sec:methods}). The best fit is defined as the {1D weighted median of the distribution}. The combined endpoint is calculated using the correlated mean of the 12 ring-wise endpoints per sample.} 
    \label{fig:result}
\end{figure} 

\begin{table*}[]
        \centering
        \caption{Breakdown of uncertainties of the neutrino-mass-squared best fit, sorted by size. The uncertainties of the individual systematic effects are quoted in the main text.}
        \label{tab:systematics_breakdown-final}
        \begin{tabular}{lc}
            \toprule
            Effect & \SI{68.2}{\percent} CL uncertainty on \mnutwo\  (\SI{}{\electronvolt\squared})\\
            \midrule
            Statistical	& 0.29 \\
            \midrule
            {Non-Poissonian background}  & {0.11}  \\
			Source-potential variations & {0.09}\\
			Scan-step-duration-dependent {background} & 0.07 \\
			$qU$-dependent {background} & 0.06    \\
            Magnetic fields & 0.04  \\
			Molecular final-state distribution & 0.02  \\ 
			Column density $\times$ {inelastic scat.} cross-section & 0.01  \\
			Activity fluctuations & 0.01   \\
		    Energy-loss function	& $<0.01$   \\
			Detector efficiency & $<0.01$  \\
			Theoretical corrections & $<0.01$\\
			{High voltage stability and reproducibility} & {$<0.01$}\\
		\midrule
			\textbf{Total uncertainty} & 0.34   \\
        \bottomrule
       \end{tabular}
    \end{table*}

\section{Results and discussion}
The $\chi^2$ minimisation reveals an excellent goodness of fit with a $\chi^2$ per degree of freedom of $0.9$, corresponding to a p-value of 0.8. For the best fit of the squared neutrino mass we find $\mnutwo = 0.26^{+0.34}_{-0.34}\SI{}{\electronvolt\squared}$, see Fig.\ \ref{fig:result}. The independent analysis methods agree within about \SI{5}{\percent} percent of the total uncertainty. The total uncertainty on the fit is dominated by the statistical error followed by uncertainties of background parameters and the source electric potential. The full breakdown of uncertainties can be found in Tab.\ \ref{tab:systematics_breakdown-final}.

Based on the best-fit result we obtain an upper limit of $\mnu < \SI{0.9}{\electronvolt}$ at \SI{90}{\percent} CL, using the Lokhov-Tkachov method~\cite{Lokhov:2015zna}. The Feldman-Cousins technique~\cite{Feldman:1997qc} yields the same limit for the obtained best fit. This result is slightly higher than the sensitivity of $\SI{0.7}{\electronvolt}$, due to the positive fit value, which is consistent with a $\approx0.8\sigma$ statistical fluctuation assuming a true neutrino mass of zero. We also perform a Bayesian analysis of the data set, with a positive flat prior on \mnutwo. The resulting Bayesian limit at \SI{90}{\percent} credibility is {$\mnu < \SI{0.85}{\electronvolt}$}.

The ring-averaged, fitted effective endpoint is $E_{0}=\SI{18573.69\pm0.03}{\electronvolt}$. Taking into account the center-of-mass molecular recoil of
$\ce{T2}$ ($\SI{1.72}{\electronvolt}$), as well as the absolute electric source potential $\phi_{\mathrm{WGTS}}$ ($\sigma(\phi_{\mathrm{WGTS}})=\SI{0.6}{\volt}$) and the work function of the main spectrometer $\phi_{\mathrm{MS}}$ ($\sigma(\phi_{\mathrm{MS}})=\SI{0.06}{\electronvolt}$), we find a $Q$-value of $\SI{18575.2\pm0.6}{\electronvolt}$, which is consistent with the previous KATRIN neutrino-mass campaign ($\SI{18575.2\pm0.5}{\electronvolt}$ \cite{Aker:2019qfn}) and the calculated $Q$-value from the
$\ce{^3He}-\ce{^3H}$ atomic mass difference of $\SI{18575.72\pm0.07}{\electronvolt}$~\cite{Myers2015}. The good agreement underlines the stability and accuracy of the absolute energy scale of the apparatus.

We combined the neutrino-mass results from this work with the previously published KATRIN (KNM1) result~\cite{Aker:2019qfn}. {A simultaneous fit of both data sets yields $\mnutwo = \SI{0.1\pm0.3}{\electronvolt\squared}$ and a corresponding upper limit of $\mnu<\SI{0.8}{\electronvolt}$ at \SI{90}{\percent} CL, based on Lokhov-Tkachov or the Feldman-Cousins technique. The same result is obtained when multiplying the \mnutwo\ distributions from Monte-Carlo propagation, or adding the $\chi^2$~profile of the individual fits. As both data sets are statistics-dominated, correlated systematic uncertainties between both campaigns are negligible. Furthermore, we investigated a Bayesian combination, where the KNM1 posterior distribution of \mnutwo\ is used as prior for the KNM2 analysis, yielding consistent results. More details on the combined analyses can be found in the supplementary material.}





\begin{figure}
    \centering
    \includegraphics[width=\linewidth]{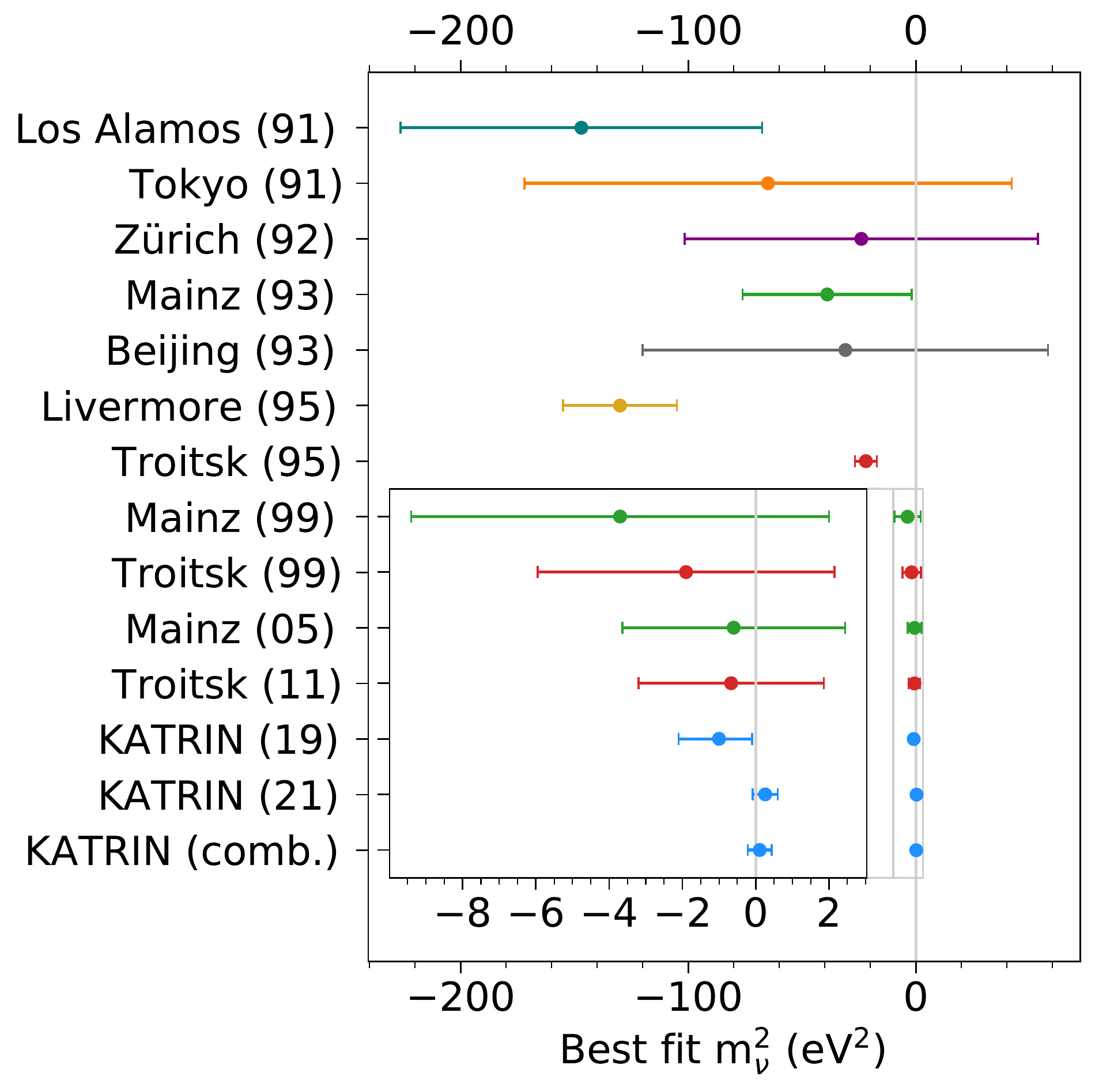}
    \caption{Comparison of {best-fit} values and uncertainties with previous neutrino mass experiments. References: Los Alamos 91 \cite{Robertson:1991vn}, Tokyo 91 \cite{Tokyo1991}, Zürich 92 \cite{Zurich1992}, Mainz 93 \cite{Mainz1993}, Beijing 93 \cite{beijing:1993}, Livermore 95 \cite{Livermore1995}, Troitsk 95 \cite{Troitsk1995}, Mainz 99 \cite{Mainz1999}, Troitsk 99 \cite{Troitsk1999}, Mainz 05 \cite{Kraus2005}, Troitsk 11 \cite{Aseev2011}, KATRIN 19 \cite{Aker:2019uuj, aker2021analysis}, KATRIN 21: this work, KATRIN combined: KATRIN 19 combined with KATRIN 21. }
    \label{fig:comparison}
\end{figure}

\section{Conclusion and Outlook}
The second neutrino-mass measurement campaign of KATRIN, presented here, is the first direct neutrino-mass measurement reaching sub-eV sensitivity (\SI{0.7}{\electronvolt} at \SI{90}{\percent} CL). Combined with the first campaign we set an improved upper limit of $\mnu<\SI{0.8}{\electronvolt}$ (\SI{90}{\percent} CL). We therefore have narrowed down the allowed range of quasi-degenerate neutrino-mass models and we have provided model-independent information about the neutrino mass, which allows the testing of non-standard cosmological models~\cite{Esteban:2021ozz, workgroup2020cosmobit}. Fig.\ \ref{fig:comparison} displays the evolution of best-fit \mnutwo\ results from historical neutrino-mass measurements up to the present day. 


Compared to its previous measurement campaign, the KATRIN experiment has decreased the statistical and systematic uncertainties by about a factor of three and two, respectively. With the total planned measurement time of 1000 days, the total statistics of KATRIN will be increased by another factor of 50. A reduction of the background rate by a factor of two will be achieved by an optimised electromagnetic field design of the spectrometer section. Moreover, by eliminating the radon- and Penning-trap-induced background, the background-related systematic effects are expected to be significantly reduced. Together with a new high-rate krypton calibration scheme and a new method to improve the determination of the magnetic fields we will minimise the dominant systematic uncertainties to reach the target sensitivity of \mnutwo\ in the vicinity of $\SI{0.2}{\electronvolt}$. 

High-precision $\upbeta$-spectroscopy with the KATRIN experiment has proven to be a powerful means to probe the neutrino mass with unprecedented sensitivity and to explore physics beyond the Standard Model, such as sterile neutrinos~\cite{Aker:2020vrf}. Together with cosmological probes and searches for neutrinoless double $\upbeta$-decay\cite{Dolinksi2019}, the upcoming KATRIN data will play a key role in measuring the neutrino mass parameters.

\bibliographystyle{naturemag-doi}
\bibliography{knm2-bib}

\section{Methods}
\label{sec:methods}
In this section we describe the data analysis chain starting from the data processing to the high-level fit and limit setting. Moreover, we provide details on one of the key calibration campaigns, concerning the source electric potential. For a more extensive description of the KATRIN analysis procedure, the reader is referred to a recent publication~\cite{aker2021analysis}.

\subsection{Data processing, selection, and combination}
The first step of the analysis chain is the preparation of the data. Raw data are combined into integral spectral data points, which are then fitted with an analytical spectrum prediction including the response of the experiment.

\paragraph{Rate determination}

The electrons which are transmitted through the main spectrometer are further accelerated by a post-acceleration electrode (PAE) with the potential $U_{\mathrm{PAE}}=\SI{10}{\kilo\electronvolt}$ before they are detected by the focal-plane detector~\cite{Amsbaugh:2014uca}. The latter provides a high detection efficiency (>\SI{95}{\percent}) and a moderate energy resolution (\SI{2.8}{\kilo\electronvolt} FWHM at \SI{28}{\kilo\electronvolt}). The total rate per pixel at a given retarding potential $qU_{i}$ is determined by integrating the rate in a wide and asymmetric region of interest (ROI) of \mbox{$\SI{14}{\kilo\electronvolt} \leq E \leq \SI{32}{\kilo\electronvolt}$}. The asymmetric ROI is chosen to account for energy losses in the deadlayer of the Si-PIN diodes~\cite{Wall2014}, for partial energy deposition due to backscattering of electrons off the detector surface and due to charge sharing between pixels. 

The detector efficiency slightly depends on $qU_{i}$. Three effects are considered: 1) The differential energy spectrum of the transmitted electrons at the FPD is shifted (and slightly scaled) according to $qU_{i}$. Since the same ROI is used for each $qU_{i}$, some electrons are no longer covered by the fixed ROI when lowering the potential, which effectively changes the detection efficiency. Based on reference measurements the relative reduction of detection efficiency at $E_0-\SI{1}{\kilo\electronvolt}$, with respect to the efficiency at $E_0$, is determined to be $\delta_{\mathrm{ROI}} = 0.002$, with an uncertainty of $\SI{0.16}{\percent}$. The data is corrected according to the efficiency at the given $qU_{i}$ value. 2) As the counting rate at the focal plane detector depends on $qU_{i}$, so does the probability of pile-up. As the energy of most pile-up events is added, they are not covered by the ROI, thereby effectively changing the detector efficiency with $qU_{i}$. Based on a random-coincidence model assuming a Poisson-distributed signal, the relative reduction of efficiency at $E_0-\SI{1}{\kilo\electronvolt}$, with respect to the efficiency at $E_0$, is estimated to be $\delta_{\mathrm{PU}} = 0.0002$, with an uncertainty of $\SI{18}{\percent}$. The data is corrected accordingly. 3) Finally, a significant fraction of about \SI{20}{\percent} of all electrons impinging on the detector surface are backscattered. For low retarding potentials and small energy depositions in the detector, these backscattered electrons have a chance of remaining undetected by overcoming the retarding potential a second time in the direction towards the tritium source. The lower $qU_{i}$, the higher is the probability for lost electrons, effectively changing the detection efficiency. Based on Monte Carlo simulations, we estimate the efficiency reduction to be $\delta_{\mathrm{BS}} < 0.001$ at $E_0-\SI{1}{\kilo\electronvolt}$. We neglect this effect in this analysis. Systematic uncertainties related to the efficiency corrections are negligibly small for the presented analysis, which only considers the last $\SI{40}{\electronvolt}$ of the tritium spectrum.

\paragraph{Selection of golden pixels and scans}
The detector is segmented into 148 pixels of equal area. For the analysis 117 pixels have been chosen. The rejected pixels show undesirable characteristics such as broadened energy resolution, higher noise levels, or decreased rate due to misalignment of the beam-line with respect to the magnetic flux tube. 

Out of the 397 scans recorded, 361 passed a strict quality assessment and were included in the neutrino mass analysis. The other runs were rejected for reasons of failed set-points of spectrometer electrodes and downtime of the Laser Raman system. The golden pixel and run selection is performed prior to the unblinding of the data.

\paragraph{Data combination}
The 117 golden pixels are grouped into 12 detector rings and the detector counts recorded within each ring are summed to obtain 12 independent spectra. All golden scans are combined by adding the counts recorded at the same retarding energies. This method leads to a total number of data points of $n_{\mathrm{data}} = n_{qU}\times n_{\mathrm{rings}} = 336$.

\subsection{Spectrum calculation}
In order to infer the neutrino mass, the integral spectrum is fitted with the analytical spectrum calculation including the experimental response. As can be seen in Eq.~(\ref{Ntheo}), the theoretical spectrum is composed of the differential tritium spectrum $R_\upbeta(E)$ and the experimental response function $f(E, qU_i)$. The differential spectrum is given by

\begin{align}
\begin{split}
R_{\mathrm{\upbeta}}(E)
& =
\frac{\CouplingSq{F} \cos^2\cabibbo }{2 \pi^3} |M_{\mathrm{nucl}}|^2 F(E, Z' =2) \\
&
\cdot \phaseSpace{(E+m_e)}{m_e} \\
&
\cdot \sum_{f}
\zeta_f \,
\phaseSpace{\neutrinoEnergy(E)}{m_\nu}
\heavyside(\neutrinoEnergy(E) - m_\nu)
.
\end{split}
\label{eq:tritium_spec_diff}
\end{align}

with $\Coupling{F}$ denoting the Fermi constant, $\cos^2\cabibbo$ the Cabibbo angle, $|M_{\mathrm{nucl}}|^2$ the energy-independent nuclear matrix element, and $F(E, Z' =2)$ the Fermi function. While,  $\neutrinoEnergy(E) = \reducedEndpoint - V_f - E$, where $\reducedEndpoint$ denotes the maximum kinetic energy of the electron, in case of zero neutrino mass, and $V_f$ describe the molecular excitation energies, which are populated with the probabilities $\zeta_f$. $E$ and $m_{\mathrm{e}}$ denote the kinetic energy and mass of the $\upbeta$-electron, respectively.

Beyond the molecular effects, further theoretical corrections arise on the atomic and nuclear level~\cite{Mertens:2014nha}. Relevant for this analysis are only the radiative corrections to the differential spectrum, which are included in the analytical description, but not shown here. Furthermore, we emulate the effect of Doppler broadening as well as spatial and temporal source and spectrometer electric potential variations, by broadening the final state distribution with a Gaussian distribution.

\begin{figure}
    \centering
    \includegraphics[width=\linewidth]{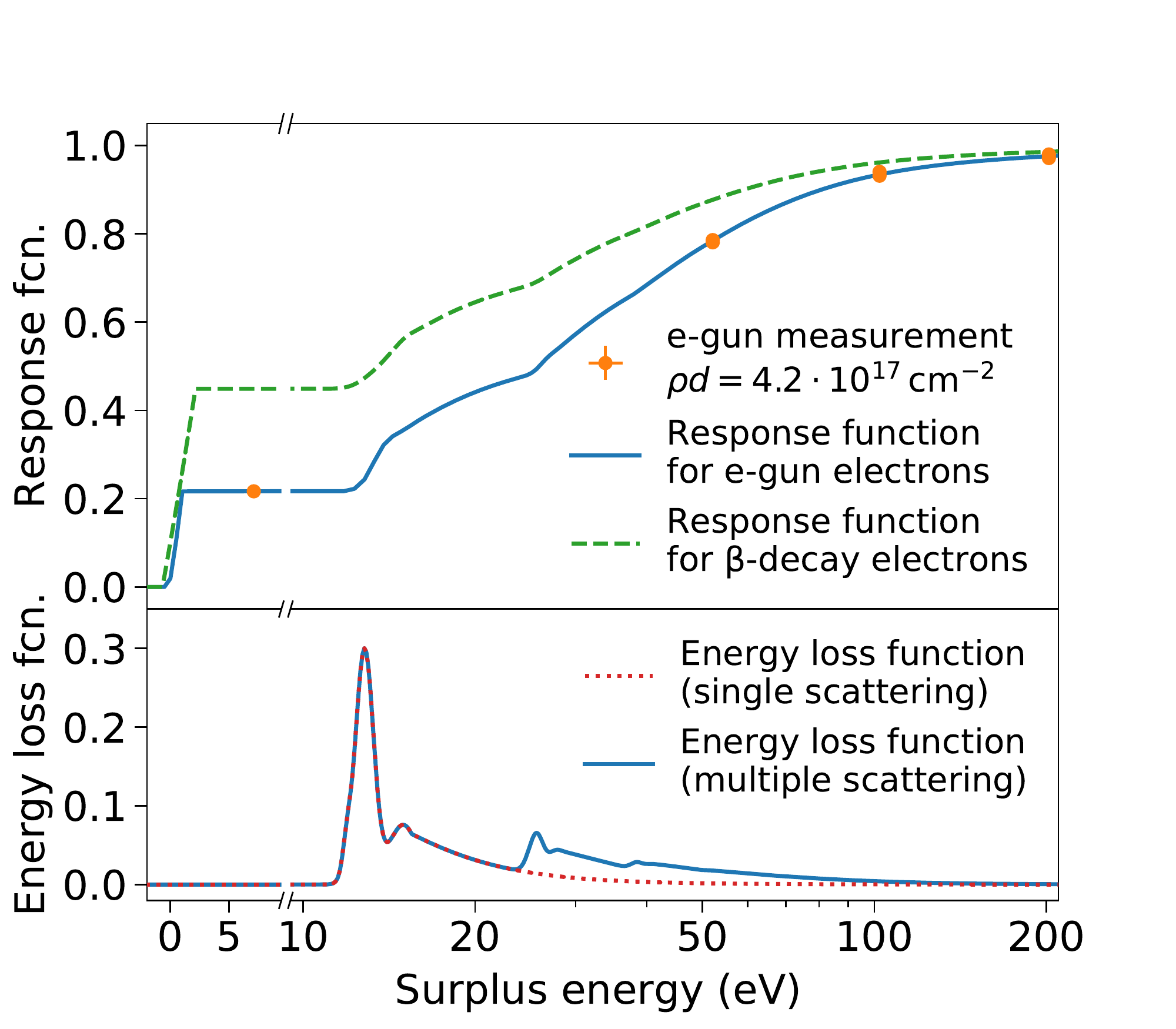}
    \caption{(Top) Electron-gun rate {from multiple measurements} at four surplus energies (yellow circles) and fitted response function of the electron-gun electrons (blue line) which start at the rear of the source for the best-fit column density of $\rho d=\SI{4.2e17}{\per\centi\metre\squared}$ for that specific measurement. The green-dashed line illustrates the response function of the beta-decay electrons starting inside of the gaseous tritium source for the average column density during the neutrino mass measurement campaign of $\rho d=\SI{4.2e17}{\per\centi\metre\squared}$. (Bottom) Energy-loss function for a single scattering process (red-dotted line) and multiple scatterings (blue line) as they occur for the measured $\rho d$ value.}
    \label{fig:eloss-rhod}
\end{figure}

The experimental response function
\begin{align}
\begin{split}
f(E-qU)
& = \int_{\epsilon=0}^{E-qU} \! \int_{\theta=0}^{\theta_\mathrm{max}} \mathcal{T}(E-\epsilon, \theta, U) \sin\theta \\
&\cdot \sum_s P_s(\theta) \, f_s(\epsilon) \, d\theta\, d\epsilon ~ .
\end{split}
\label{eq:transm_func_det}
\end{align}


depicts the probability of an electron with a starting energy $E$ to reach the detector. It combines the transmission function $\mathcal{T}(E-\epsilon, \theta, U)$ of the main spectrometer and the electron's energy losses $\epsilon$ due to inelastic scattering with the tritium molecules in the source. The scattering energy losses are described by the product of the $s$-fold scattering probabilities $P_s(\theta)$, which depend on the path length through the source and hence on the pitch angle $\theta$, and the energy-loss function $f_s(\epsilon)$ for a given number of scatterings $s$. The energy-loss function is determined experimentally with the electron gun installed at the rear of the source. A typical response function and corresponding energy loss function is shown in Fig.\ \ref{fig:eloss-rhod}.


The integrated transmission function for an isotropic source of electrons is given by

\begin{align}
T&(E,U) = \int_{\theta=0}^{\theta_\text{max}} \; \mathcal{T}(E,\theta,U) \cdot \sin \theta \, d\theta \\
& =
\left\{\begin{array}{ll}
0 &, \epsilon<0 \\
1 - \sqrt{1-\frac{E-qU}{E} \frac{B_\text{source}}{B_\mathrm{ana}} \frac{2}{\gamma\!+\!1}} \, &, 0\leq E-qU \leq \Delta E \\
1 - \sqrt{1-\frac{B_\text{source}}{B_\text{max}}} &, E-qU >\Delta E
\end{array}
\right. \, .
\label{eq:response:tf_simple}
\end{align}

It is governed by the magnetic fields at the starting position $B_\text{source}$ of the electron, the maximum field $B_\text{max}$ in the beam line, and the magnetic field in the spectrometer's analysing plane $B_\mathrm{ana}$. Synchrotron
energy losses of $\upbeta$-electrons in the high magnetic field
in the source and transport systems are included as an
analytical correction to the transmission function (not shown here).

\begin{table*}[]
    \centering
    \caption{Overview of key figures for {the} second neutrino mass {measurement} campaign. When value and uncertainty are both given in \%, then the statement is understood as $\left(\mathrm{value}\pm\mathrm{uncertainty}\right)\,\%$. $^*$The analysis interval refers to $qU_i\in \left(E_{0} - \SI{40}{\electronvolt}, E_{0} + \SI{135}{\electronvolt}\right)$.} 
    \begin{tabular}{lll}
    \toprule
          &  Value & Uncertainty\\
        \midrule
        \multicolumn{3}{c}{Average source parameters}\\
        Column density                              & $\SI{4.23e17}{\per\centi\metre\squared}$ & $\SI{0.25}{\percent}$\\
        Source activity                             & $\SI{9.5e10}{\becquerel}$                     & {-}\\
        {Relative activity variation within scan} & $\SI{3.5e-4}{}$                     & {-}\\
        {Atomic} tritium purity                       & {0.987}                              &  {0.003}\\
        Concentration T$_2$                         & {0.973}                              & {0.005}\\
        Concentration DT & {0.0031}                               & {0.0004}\\
        Concentration HT    & {0.023}                               & {0.004}\\
        Source temperature                          & $\SI{30.065}{\kelvin}$                & {$\SI{0.001}{\kelvin}$}    \\
        \multicolumn{3}{c}{Final state distribution}\\
        {Probability to decay to the ground state}           & \SI{57}{\percent} & \SI{1}{\percent}\\
        {Variance of ground state distribution}
                   & {\SI{0.2}{\electronvolt\squared} }    & {\SI{0.5}{\percent}}\\
        {Variance of FSD in analysis interval$^*$}             & {\SI{185.2}{\electronvolt\squared}}       & {\SI{1.2}{\percent}}\\ 
        \multicolumn{3}{c}{Source plasma parameters}\\
        Variance $\sigma_{\mathrm{P}}^2$            & $\SI{12.4e-3}{\electronvolt\squared}$ & $\SI{16.1e-3}{\electronvolt\squared}$\\
        {Rear-to-front} asymmetry parameter $\Delta_{\mathrm{P}}$   & $\SI{0} {\milli\electronvolt}$        & $\SI{61}{\milli\electronvolt}$\\
        \multicolumn{3}{c}{Average magnetic fields}\\
        Source                                      & $\SI{2.52}{\tesla}$                   & \SI{1.7}{\percent}\\
        Analysing plane                              & $\SI{6.308e-4}{\tesla}$    & \SI{1.0}{\percent}\\
        Maximum                                     & $\SI{4.239}{\tesla}$                  & \SI{0.1}{\percent}\\
        \multicolumn{3}{c}{Spectrometer parameters}\\
        Background                                  &  $\SI{220}{mcps}$                     & $\SI{0.8}{mcps}$ \\
        Background increase during scan step        &  $\SI{3}{\micro cps\per\second}$      & $\SI{3}{\micro cps\per\second}$ \\
        Background qU {linear} dependence                    & $0\,\mathrm{mpcs/keV}$                & $4.74\,\mathrm{mpcs/keV}$  \\
        Non-Poissonian background                   & \SI{11.2}{\percent}                               & -\\
        High voltage stability and reproducibility                & \SI{13.5}{meV}                        &  \SI{0.8}{meV} \\
        \multicolumn{3}{c}{Detector parameters}\\
        Detector efficiency                         & {$>\SI{95}{\percent}$} & -\\
        Region of Interest (ROI)                    & $14-\SI{32}{\kilo\electronvolt}$ & -\\
        Detector efficiency qU {reduction} (ROI)     & {$\SI{2e-3}{}$} (\SI{1}{\kilo\electronvolt} below $E_0$)     &  \SI{0.16}{\percent} \\
        Detector efficiency qU {reduction} (pile up) & {$\SI{2e-4}{}$} (\SI{1}{\kilo\electronvolt} below $E_0$)     &  \SI{18}{\percent} \\
         \bottomrule
    \end{tabular}
    \label{tab:key-figures}
\end{table*}

\subsection{Systematic uncertainties}
The analytical description $R(qU_i, r_j)$ of the integral $\upbeta$-spectrum, shown in Eq.~(\ref{Ntheo}), contains various signal- and background-related parameters, which are known with a certain accuracy. In the following we describe these parameters, their uncertainties, and their treatment in the neutrino mass analysis. 

\paragraph{Signal-related systematic effects}
The spectrum prediction includes uncertainties of the nine parameters of the empirical energy loss function (the individual relative uncertainties of the parameters $\sigma_{\mathrm{eloss},k}$ are between $\SI{0.016}{\percent}$ and $\SI{3.8}{\percent}$), a relative uncertainty of the product of column density and scattering cross-section ($\sigma_{\rho d \times \sigma}=\SI{0.25}{\percent}$), relative uncertainties of the theoretical description of the molecular final-state distribution ($\sigma_{\mathrm{FSD}}=\mathcal{O}(\SI{1}{\percent})$), and relative uncertainties of the magnetic field in the source ($\sigma_{B_{\mathrm{source}}}=\SI{1.7}{\percent}$), {in} the analysing plane ($\sigma_{B_{\mathrm{ana}}}=\SI{1}{\percent}$), and the maximal field ($\sigma_{B_{\mathrm{max}}}=\SI{0.1}{\percent}$). The variations of the {$\upbeta$}-decay activity during a scan ($\sigma_{\mathrm{scan}}=\SI{0.03}{\percent}$) were negligibly small.

The spatial and temporal variations of the source electric potential were not included in the first neutrino mass campaign. With the increase of the source column density from \SI{1.11e17}{\per\centi\metre\squared} to \SI{4.23e17}{\per\centi\metre\squared} in this campaign, the creation rates and densities of the charge carriers, as well as the fraction of scattered electrons, increased accordingly, which makes plasma effects more relevant~\cite{KDR2004,KleinPhD2018,Nastoyashchii:FST2005}. A detailed description of the plasma calibration is given in Sec.\ \ref{ssec:sourcepotential}.


\paragraph{Background-related systematic effects} Electrons created in a radon decay in the main spectrometer volume can initially be magnetically trapped. These trapped electrons can create a cluster of 10 - 100 secondary electrons~\cite{Mertens:2012vs} by scattering off the residual gas in the main spectrometer, which is operated at a pressure of $10^{-11}$~mbar. These secondary electrons arrive at the detector within a time window of about $\SI{1000}{\second}$, hence leading to a non-Poissonian rate distribution. The observed background rate can be modelled by a Gaussian distribution, with a width $\SI{11.2}{\percent}$ wider than expected from a purely Poisson distribution. This overdispersion is treated as an increased statistical uncertainty in the analysis.

As the transmission conditions for the background electrons slightly depend on the retarding-potential setting, a small retarding-potential dependence of the background can occur. In the analysis, we allow for a linear dependence of the background on the retarding potential and use dedicated test measurements to constrain the possible slope to $m^{qU}_{\mathrm{bg}} = \SI{0 \pm 4.74}{mcps\per\kilo\electronvolt}$.
    
Finally, the pre- and main spectrometers, being both operated at high voltage, create a Penning trap between them. Stored electrons and the subsequently produced positive ions, which can escape the trap into the main spectrometer, are a source of background, as illustrated in Fig.\ \ref{fig:experiment}. To mitigate this background, the trap is emptied with an electron-catcher system~\cite{Aker:2019nlx} after each scan step. However, a potentially small increase of the background rate within a scan step cannot be excluded, which could lead to a background dependence on the duration of the scan step. By fitting a linear increase to the rate evolution within all scan steps we find a slope of $m^{\mathrm{t}_{\mathrm{scan}}}_{\mathrm{bg}} = \SI{3 \pm 3}{\micro cps\per\second}$, which is included in the {$\upbeta$-spectrum} fit~\footnote{This effect was first observed in later measurement campaigns, in which the scan-step duration was significantly increased. It was therefore only included after the data was already un-blinded to a sub-group of the collaboration. As the evaluation of the size of this uncertainty was provided by an independent task group of the collaboration, the reported result remains bias-free.}.

\subsection{Source Potential Calibration }
\label{ssec:sourcepotential}

The absolute electric potential of the source does not affect the spectral shape of the measured spectrum. An unknown absolute source potential is largely absorbed by the effective endpoint, which is a free parameter in the fit. A change of the effective endpoint mostly leads to a shift of the spectrum and has a negligible effect on the spectral shape. Accordingly, an unknown radial variation of the electric potential is in good approximation absorbed by the ring-wise endpoint parameters. Consequently, these effect have a minor impact on the neutrino mass analysis. However, any temporal or longitudinal variations of the source potential can lead to spectral distortions, which are parameterised by a Gaussian broadening $\sigma_{\mathrm{P}}$ and 
the parameter $\Delta_{\mathrm{P}}$ which quantifies the longitudinal rear-to-front asymmetry of the electric potential of the source. This asymmetry of the potential results in a shift of the energy spectrum associated with scattered electrons (predominantly originating from the rear of the source tube) compared to the spectrum of the unscattered electrons (predominantly originating from the front of the source tube).

Both parameters are assessed with the help of co-circulating $^{83\mathrm{m}}\mathrm{Kr}$ gas. The spectroscopy of its mono-energetic conversion electron lines reveals information about the broadening $\sigma_{\mathrm{P}}$ of the line shape, from which an upper limit of $\Delta_{\mathrm{P}}$ is derived.

The calibration was performed at an elevated source temperature of $T=\SI{80}{\kelvin}$ to prevent condensation of the Kr gas (as compared to the $T=\SI{30}{\kelvin}$ set-point used during neutrino-mass measurements). The tritium circulation loop of the tritium source \cite{STURM2021} can be operated in two modes, a) in a mode with direct-recycling of the krypton-tritium mixture which is limited to a maximum column density of $\SI{40}{\percent}$ ($\SI{2.08e17}{\per\centi\metre\squared}$), but delivers a high krypton rate, and b) in a mode of fractional direct-recycling which can be operated up to $\SI{75}{\percent}$ ($\SI{3.75e17}{\per\centi\metre\squared}$) of the nominal column density, but with only about $\SI{0.5}{\percent}$ of the maximum krypton activity as compared to mode a). 

\begin{figure}
    \centering
    \includegraphics[width=\columnwidth]{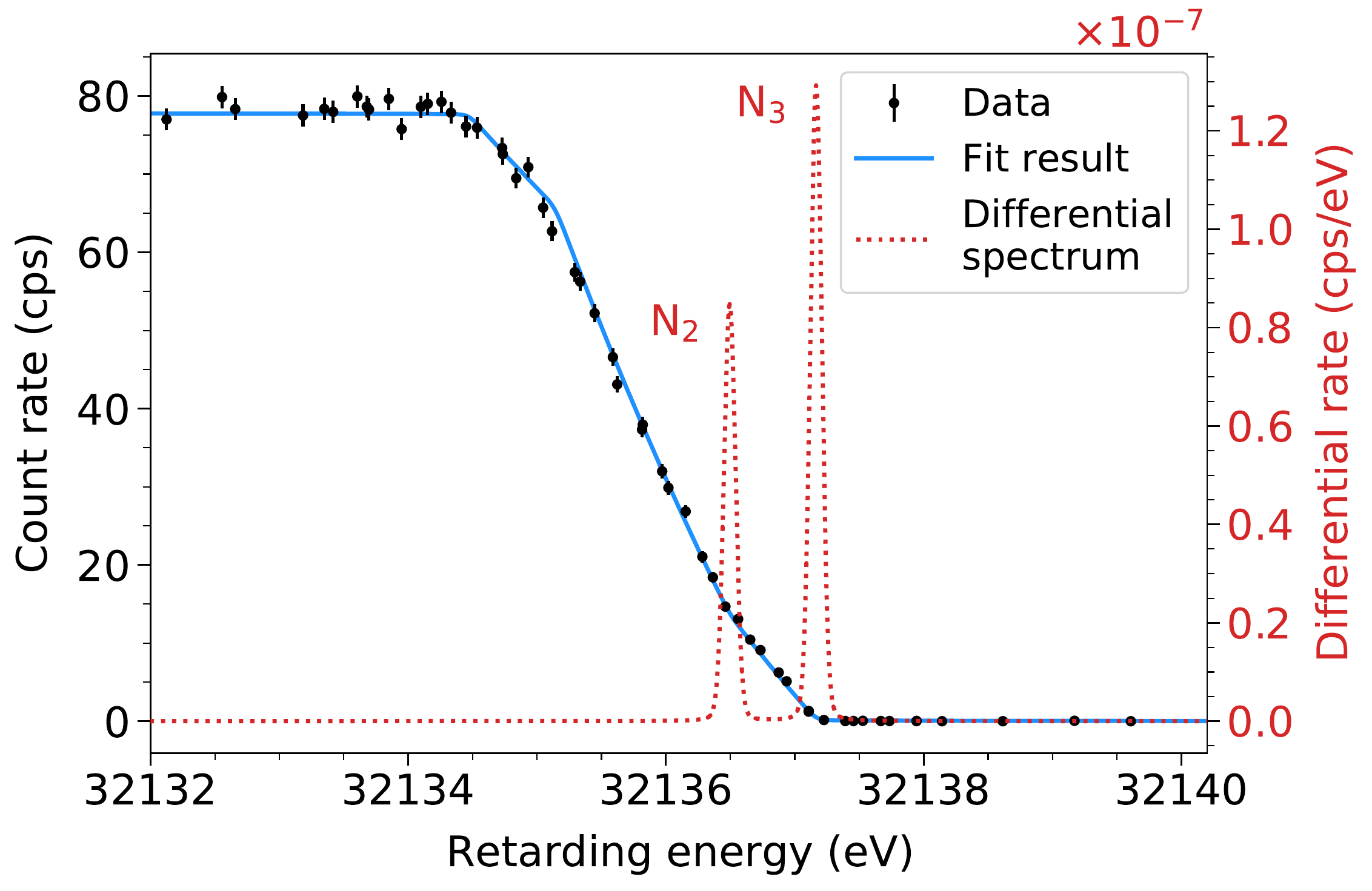}
    \caption{$N_{2}$/$N_{3}$ doublet (internal conversion lines of $^{83\mathrm{m}}$Kr) at $\SI{40}{\percent}$ of the nominal column density for a {sample data} spectrum. {The visible width of the integrated spectrum is dominated be the energy resolution}. The differential line width represents the broadening by {fitted} $\sigma_{\mathrm{P}}^2$ only. The Doppler line width from {the} thermal movement of the Kr atoms is not displayed. {The goodness-of-fit is $\chi^2/\mathrm{ndof} = 68.2 / 52 = 1.3$ and the p-value is $6.6\times10^{-2}$.}}
   \label{fig:plasma}
\end{figure}

The plasma parameters are inferred by combining the measurements of the internal conversion lines N$_{2,3}$-32 and L$_3$-32. The energy of the conversion electrons, emitted from a particular subshell of the $^{83\mathrm{m}}\mathrm{Kr}$ atom, is \SI{30472.6}{\electronvolt} for the L$_3$-32 singlet and \SI{32137.1}{\electronvolt} (\SI{32137.8}{\electronvolt}) for the N$_{2,3}$-32 doublet~\cite{Altenmueller:JPG2020}. The L$_3$-32-line, on the one hand, has a high intensity, but its natural line width is not known precisely~\cite{Venos:2018tnw}. The N$_{2,3}$-32-lines, on the other hand, have a low intensity, but their natural line width is negligible {compared to a spectral broadening caused by variations of the electric potential.} Thus, any broadening of the N$_{2,3}$-32-lines beyond the known spectrometer resolution and the thermal Doppler broadening can be assigned to variations of the electric potential within the source. Based on the N$_{2,3}$-32-line doublet measurement at \SI{40}{\percent} of the nominal column density we find $\sigma_{\mathrm{P}}^2(\SI{40}{\percent})=\left(1.4\pm0.3\right)10^{-3}\SI{}{\electronvolt\squared}$ (See Fig.\ \ref{fig:plasma}). Combined with L$_3$-32-line measurements both at \SI{40}{\percent} and \SI{75}{\percent} {of the nominal} column density, we assess the relative change of the broadening and find $\sigma_{\mathrm{P}}^2(\SI{75}{\percent})=\left(8.0\pm8.2\right)10^{-3}\SI{}{\electronvolt\squared}$. Finally, we conservatively extrapolate {with exponential scaling} this value to \SI{84}{\percent} {of the nominal} column density, leading to $\sigma_{\mathrm{P}}^2(\SI{84}{\percent})=\left(12.4\pm16.1\right)10^{-3}\SI{}{\electronvolt\squared}$. 
\footnote{{
The plasma parameters and uncertainties based on the calibration measurements at $\SI{80}{\kelvin}$ cover the conditions during the neutrino mass measurements at $\SI{30}{\kelvin}$. The effect by the different temperatures is smaller than the assumed uncertainty. For future campaigns the tritium source is operated at $\SI{80}{\kelvin}.$}}  


{In the analysis, we limit the broadening $\sigma_{\mathrm{P}}$ to positive values. We construct the probability density function of the asymmetry parameter $\Delta_{\mathrm{P}}$ based on {phenomenological} considerations on the relation of both plasma parameters given by $|\Delta_{\text{P}}| \leq \sigma_{\mathrm{P}}/{1.3}$ ~\cite{Machatschek:2020}. For many samples of $\sigma_{\mathrm{P}}^2$ we draw a value for $\Delta_{\mathrm{P}}$ from a uniform distribution in the range $-\sigma_{\mathrm{P}}/1.3 \leq \Delta_{\mathrm{P}} \leq \sigma_{\mathrm{P}}/1.3$. The resulting distribution can be approximated by a Gaussian distribution centred around $\SI{0}{\milli\electronvolt}$ with a width of $\SI{61}{\milli\electronvolt}$.}

We expect to reduce the systematic uncertainties in future campaigns by operating the tritium source at the same temperature during the neutrino-mass and krypton measurements, and by using an ultra-high intensity krypton source (with about \SI{10}{\giga\becquerel} of the $^{83}$Rb mother isotope), which will make the N$_{2,3}$-32-line measurement possible at nominal column density (mode b). 

\subsection{Parameter inference}
We infer the parameters of interest via a minimisation of the $\chi^2$ function 

\begin{align}
\begin{split}
    \chi^2 = & \left(\vec{R}_\mathrm{data}(q\vec{U}, \vec{r})-\vec{R}(q\vec{U}, \vec{r}|\vec{\Theta},\vec{\eta})\right)^{T} \\
    & \cdot C^{-1}\left(\vec{R}_\mathrm{data}(q\vec{U}, \vec{r})-\vec{R}(q\vec{U}, \vec{r}|\vec{\Theta},\vec{\eta})\right), \label{eq:cov}
\end{split}
\end{align}

\label{eq:chi}
where $\vec{R}_\mathrm{data}(q\vec{U}, \vec{r})$ gives the measured count rates at a retarding potential $qU_i$ for the detector ring $r_j$, $\vec{R}(q\vec{U}, \vec{r})$ gives the predictions of these rate, and $C$ is the covariance matrix that includes the statistical uncertainties and can be used to also describe systematic uncertainties. The usage of a $\chi^2$ minimisation is justified, as the numbers of electrons per scan-step $qU_i$ and detector ring $r_j$ are sufficiently large (> 700) to be described by a Gaussian distribution, instead of a Poisson distribution. 

The fit has $1+3\times12=37$ free parameters $\vec{\Theta}$, including a single parameter for the neutrino mass squared \mnutwo, and 12 ring-wise values for each of the three spectral parameters: the normalisation factor $A_{\mathrm{s}}$, the background rate $R_{\mathrm{bg}}$, and the effective endpoint $E_0$. In addition, the spectrum depends on systematic parameters $\vec{\eta}$ (as found in Tab.\  \ref{tab:key-figures}), such as the column density, tritium isotopologue concentrations, magnetic fields, etc. These parameters are known with a certain accuracy and their uncertainty needs to be propagated to the final neutrino mass result. Four different methods are used for the KATRIN analysis:

\paragraph{Pull method} In the `pull method' the systematic {parameters} $\eta_i$ are treated as free parameters in the fit and introduced as nuisance terms in the $\chi^2$ function
\begin{align}
    \chi^2(\vec{\Theta},\vec{\eta}) &= \chi^2(\vec{\Theta},\vec{\eta}) + \sum_i\left(\frac{\hat{\eta_i}-\eta_i}{\sigma_{\eta_i}}\right)^2 ~.
\end{align}
The nuisance terms allow the parameter to vary around its best estimation $\hat{\eta_i}$ according to its uncertainty $\sigma_{\eta_i}$ as determined from external measurements 

This method is computationally intensive due to the complexity in calculating the tritium spectrum and the minimisation with respect to multiple free parameters.
For example, it is not practical to treat the uncertainties of the molecular final state distribution, which is given as a discrete list of excitation energies and corresponding probabilities, with this method. The advantage of this method is that we make the maximum use of the data. If the spectral data contain information about the systematic parameters $\vec{\eta}$, it is automatically taken into account.

\paragraph{Covariance matrix method}
As can be seen in Eq.~(\ref{eq:chi}), the standard $\chi^2$ estimator includes a covariance matrix $C$ which can describe both the statistical and systematic model uncertainties. The diagonal entries describe the uncertainties which are uncorrelated for each $R(qU_i, r_j)$, while the off-diagonal terms describe the correlated uncertainties between the $R(qU_i, r_j)$. The covariance matrix $C$ is computed by simulating $\mathcal{O}(10^4)$ $\upbeta$-spectra, with the systematic parameters $\eta_i$ varied according to their probability density functions in each spectrum. From the resulting set of spectra, the variance and covariance of the spectral points $R(qU_i, r_j)$ are determined.

As the covariance matrices for individual or combined systematic effects are computed before fitting, this method is efficient with respect to computational costs. The dimension of the covariance matrix is given by the number of data points. Therefore, the efforts for matrix calculation and inversion can be diminished by reducing the number of data points.

\paragraph{Monte-Carlo propagation method}
Generally, in the Monte-Carlo (MC) propagation technique, the uncertainties of the parameters $\eta_i$ are propagated by repeating the fit about $10^5$ times, with the systematic parameters varied according to their probability density functions each time. The method then returns the distributions of the fit parameters $\vec{\Theta}$, which reflect the uncertainty of the systematic parameters. 

More precisely, when assessing the systematic effects alone, a MC spectrum without statistical fluctuations is created (based on the best fit parameters), which is then fitted $10^5$ times with a model, which systematic parameters of interest are varied each time. Contrarily, when evaluating the statistical uncertainty alone,  $10^5$ statistically randomized MC spectra are created and fitted with a constant model. Accordingly, for obtaining the total uncertainty, both steps are combined.

To extract information on the parameters $\vec{\eta}$ from the data itself, each entry in the resulting histogram of best fit parameters is weighted with the likelihood of the corresponding fit. The final distributions are then used to estimate the best-fit values (mode of the distribution) and uncertainty (integration of the distribution up to \SI{16}{\percent} from both side).

The advantage of this method is that the number of free parameters is kept at a minimum, which facilitates the minimisation procedure. The larger number of fits, however, is time consuming and requires the usage of large computing clusters. 

\paragraph{Bayesian Inference}
In Bayesian inference one computes the posterior probability for the parameters of interest $\vec{\Theta}$ from a prior probability and a likelihood function according to Bayes' theorem. For the KATRIN analysis, the Bayesian approach has the advantage that prior knowledge on the neutrino mass can naturally be applied via a corresponding informative prior. For example, the prior of \mnutwo\ can be chosen to be flat and positive, which restricts the posterior distribution to the physically allowed regime and changes the credibility interval accordingly.

Ideally, all systematic effects $\eta_i$ would be included as free parameters constrained with our prior knowledge on the parameter. However, due to the computationally expensive spectrum calculation and the fact that the Bayesian inference requires a large number of samples in the Markov chain for sampling the posterior distribution, only the $qU$-dependent background is currently treated in this way. All other systematic uncertainties are included with a covariance matrix in the likelihood or by model variation (MV). For the MV, a large set of Markov Chains is started with randomized but fixed model variations. The randomisations are drawn from the systematic uncertainty distributions. The resulting set of posterior distributions is averaged. 

\subsection{Limit Setting}
\begin{figure}
    \centering
     \includegraphics[width=\columnwidth]{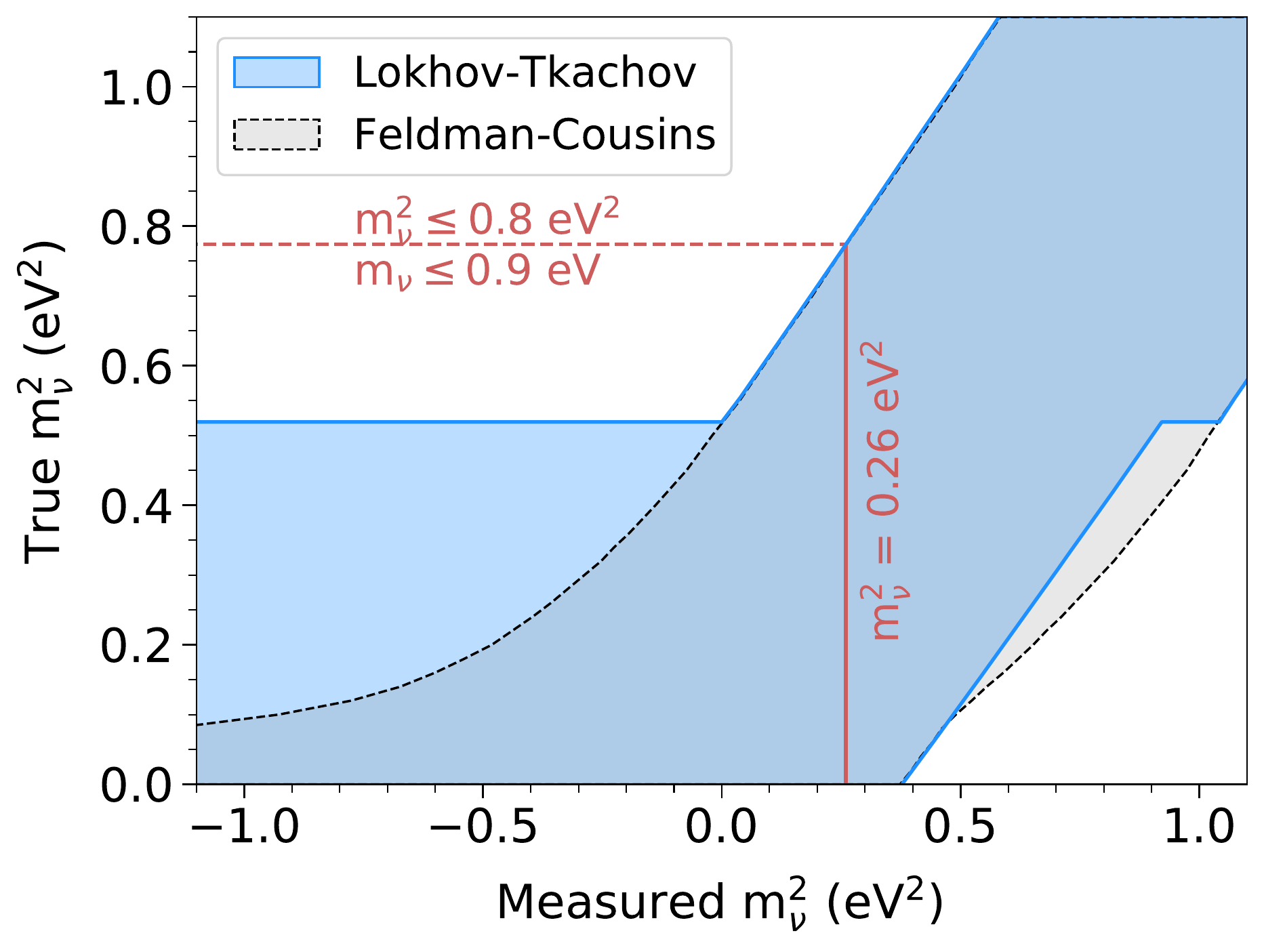}
    \caption{Frequentist bounds on the neutrino mass {for the KNM2 dataset} by using the constructions by Lokhov-Tkachov\cite{Lokhov:2015zna} and Feldman-Cousins \cite{Feldman:1997qc}.}
    \label{fig:LT-FC}
\end{figure}

We present two frequentist methods and one Bayesian method for the construction of an upper limit of the neutrino mass. For the former we use the classical Feldman-Cousins~\cite{Feldman:1997qc} and the Lokhov-Tkachov~\cite{Lokhov:2015zna} belt constructions depicted in Fig.\ \ref{fig:LT-FC}. In the Feldman-Cousins technique the acceptance region for $\widehat{m}_\nu^2$ is determined by ordering this estimator according to the likelihood ratio $\frac{\mathcal{L}\left(\widehat{m}_\nu^2\,|\,m_\nu^2\right)}{\mathcal{L}\left(\widehat{m}_\nu^2\,|\,\mathrm{max}(0,\,\widehat{m}_\nu^2)\right)}$ for a given best-fit neutrino mass squared \mnutwo. This method leads to more stringent upper limits for increasingly negative best-fit values. The method of Lokhov-Tkachov avoids this feature by using the standard Neyman belt-construction for positive values of $\widehat{m}_\nu^2$~and defining the experimental sensitivity as the upper limit in the non-physical regime of $\widehat{m}_\nu^2<0$.

\begin{figure}
    \centering
    \includegraphics[width=\columnwidth]{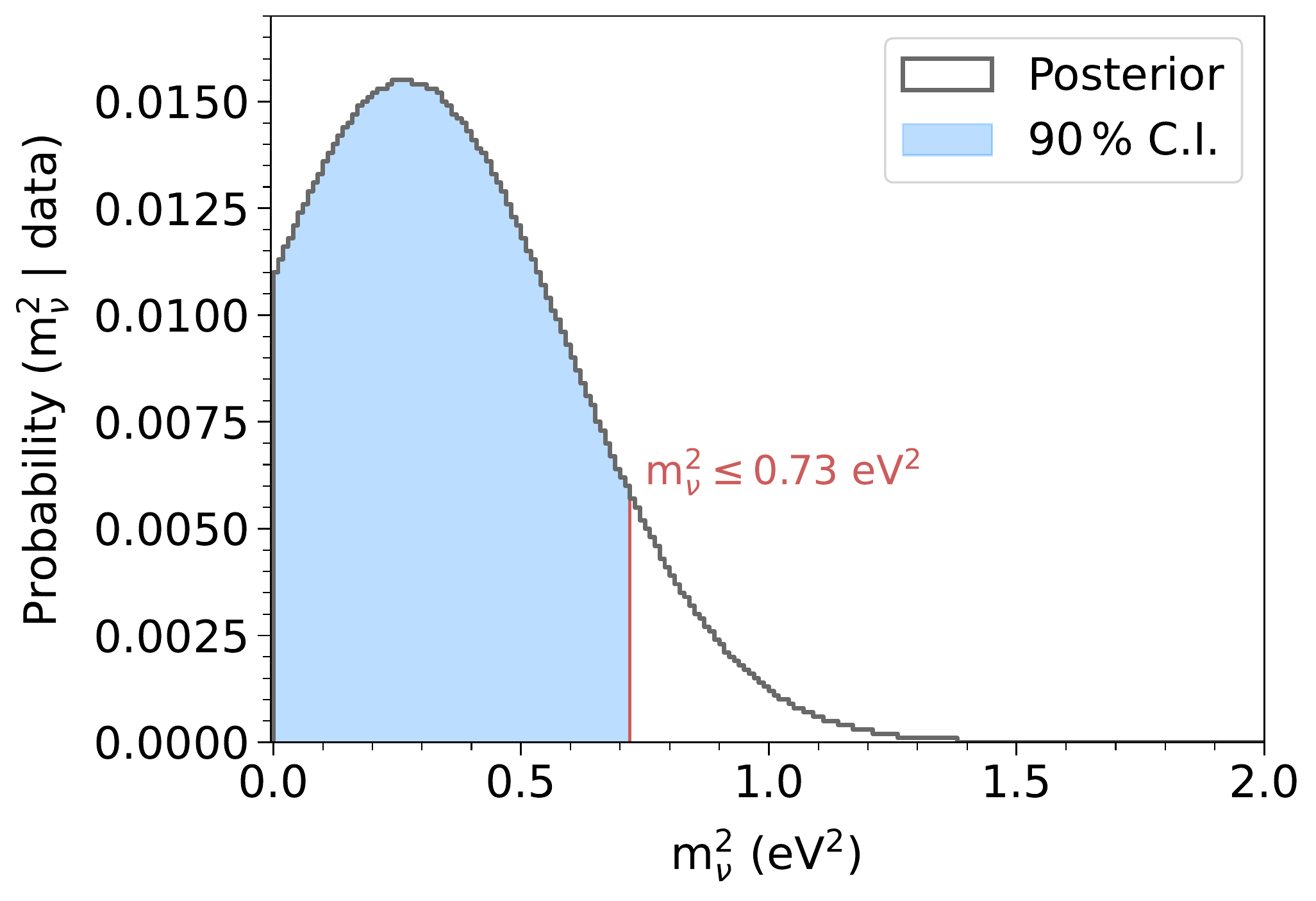}
    \caption{Posterior distribution of \mnutwo\ in the Bayesian analysis with a flat positive prior.}
    \label{fig:Bayesian}
\end{figure}

The Bayesian \SI{90}{\percent} credibility interval is obtained by integrating the posterior distribution of \mnutwo\ from zero to $\mnutwo^{\mathrm{limit}}$, such that the total probability is \SI{90}{\percent} as demonstrated in Fig.\ \ref{fig:Bayesian}. Note that the interpretation of the limit obtained in this way is different from the frequentist confidence limits and hence the numerical values may not coincide.

\subsection{Results of individual strategies}
The frequentist analyses are performed by three independent teams, which use {differing} implementations of the spectral calculation and different strategies to propagate systematic uncertainties. A Bayesian analysis is performed, which interfaces to one of the spectrum calculation softwares. The analysis by independent teams is a powerful means to cross-check individual analyses. 
The following four data analysis strategies were applied to the second data set of KATRIN.
\begin{itemize}
    \item \textbf{Strategy 1} is based on a C++ framework~\cite{KleesiekThesis} using the Minuit minimiser. It employs mainly the pull method for handling of systematics. We note that for the presented analysis, the input value for the `scan-step-duration-dependent background' was corrected after the official unblinding of the data. 
    \item \textbf{Strategy 2} is implemented in a MATLAB framework and exclusively uses the covariance matrix approach to propagate systematic uncertainties~\cite{Lisa_2020}. We note that for the presented analysis, the `scan-step-duration-dependent background' systematic was implemented only after the official unblinding of the data.
    \item \textbf{Strategy 3} is based on a C++ framework using a custom-developed minimiser~\cite{Slez_k_2020, Karl:2019pontecorvo}. In this strategy mostly the MC-propagation of uncertainties is applied. 
    \item \textbf{Strategy 4} performs a Bayesian interpretation of the data. For this approach the spectrum calculation software of `strategy 3' is interfaced with the Bayesian Analysis Toolkit (BAT)~\cite{Caldwell:2009khb}. Here, most of the systematic uncertainties are treated via the model variation (MV) technique. 
\end{itemize}
An overview of the strategies is found in Tab.\  \ref{tab:systematics_methods-all}. The resulting best fit and systematic uncertainty breakdown are listed in Tab.\ \ref{tab:systematics_breakdown-final-all}. 

\begin{table*}[h!]
        \centering
        \caption{Overview of the analysis strategies. Abbreviations: pseudo-ring: combination of three detector rings; PT = Pull term method; MC = Monte-Carlo propagation method; CM = Covariance matrix method; MV = Model variation method; param. prior = using a Bayesian prior according to parameter uncertainties. The non-Poissionan background is included as an additional statistical (stat) error.}
        \begin{tabular}{lcccc}
            \toprule
            Item & \multicolumn{4}{c}{Strategy}\\
            & 1 (Pull term) & 2 (Cov. matrix) & 3 (MC prop.)  & 4 (Bayesian)\\
            \midrule
            Minimisation & $\chi^2$ & $\chi^2$ & $\chi^2$ & posterior\\
            Detector pixel combination & 12 rings & 4 pseudo-rings & 12 rings  & 4 pseudo-rings\\
            Scan combination & stacking & stacking & stacking & stacking\\
            \midrule
            Non-Poissonian background  &  stat & stat & stat & stat \\
			Source potential variations & PT & CM & MC & MV \\
			Scan step-duration-dependent background & PT & CM & MC & MV\\
			$qU$-dependent background & PT & CM & MC & param. prior \\
            Magnetic fields & PT & CM & MC & MV \\
			Molecular final-state distribution & CM & CM & MC & CM \\ 
			Column density $\times$ cross-section & PT & CM & MC & MV \\
			Scan step fluctuations & CM & CM & CM & CM  \\
		    Energy loss 	& PT & CM & MC & MV \\
			Detector efficiency & - & CM & - & -   \\
			Theoretical corrections & - & CM & - & -    \\
			High voltage fluctuations  & - & CM & - & -    \\
            \bottomrule
       \end{tabular}
       \label{tab:systematics_methods-all}
\end{table*}

\begin{table*}[h!]
        \centering
        \caption{Breakdown of uncertainties of the neutrino mass best fit for each analysis.  Strategy 1, 2, and 4 obtain the systematic uncertainty by quadratically subtracting the statistical error from the total uncertainty, while strategy 3 assesses the systematic uncertainty directly. Strategy 4 estimates the impact of individual systematic uncertainties in fits in which all detector pixels are combined, while the other strategies use a (pseudo-)ring-dependent fit (as explained in the main text). Differences in the second digit mainly arise from numerical effects.}
        \label{tab:systematics_breakdown-final-all}
        \begin{tabular}{lcccc}
            \toprule
            & \multicolumn{4}{c}{Strategies} \\
            & 1 (Pull term) & 2 (Cov. matrix) & 3 (MC prop.) & 4 (Bayesian)\\
           \midrule
           \mnutwo\ best fit ($\SI{}{\electronvolt\squared}$) & 0.25  & 0.26 & 0.26  & 0.26 \\
        	\midrule
        	& \multicolumn{4}{c}{\SI{68.2}{\percent} CL uncertainties  ($\SI{}{\electronvolt\squared}$) }\\
        	Statistical & 0.28 & {0.28} & 0.29  &  0.31 \\
        	{Systematic} & {0.20} & {0.16} & {0.18}  &  0.14 \\
        	{Total uncertainty}
            	& 0.35 & 0.32 & 0.34   & 0.34\\
        	\midrule
        	$\chi^2$ per degree of freedom & 280.2/299 = 0.94 & 87.9/99 = 0.89 & 277.3/299 = 0.93 & - \\ 
        	\midrule
        	& \multicolumn{4}{c}{breakdown of systematic uncertainties ($\SI{}{\electronvolt\squared}$)}\\
            Non-Poissonian background  & 0.10 & {0.10} & 0.11 &  0.11\\
			Source potential variations & {0.09} & 0.07 & 0.07 & {0.07} \\
			Scan step-duration-dependent background & 0.07 & 0.07 & 0.07 & 0.04\\
			$qU$-dependent background & 0.06 & {0.05} & 0.04 & 0.06  \\
            Magnetic fields & 0.04 & {0.03} & 0.03 & 0.03\\
			Molecular final-state distribution & 0.02 & 0.01 & 0.01 & 0.01\\ 
			Column density $\times$ {inelastic scat.} cross-section & 0.01 & 0.01 & 0.01 & 0.01\\
			Scan step fluctuations & < 0.01 & < 0.01 & 0.01  & < 0.03  \\
		    Energy loss 	& < 0.01 & < 0.01 & < 0.01  & 0.02\\
			Detector efficiency & neglected & < 0.01 & neglected  &  neglected\\
			Theoretical corrections & neglected & < 0.01 & neglected  & neglected\\
			High voltage fluctuations & neglected & < 0.01 & neglected  & neglected\\
		\bottomrule
        \end{tabular}
\end{table*}

\section*{Acknowledgements}
We acknowledge the support of Helmholtz Association (HGF), Ministry for Education and Research BMBF (05A17PM3, 05A17PX3, 05A17VK2, 05A17PDA, and 05A17WO3), Helmholtz Alliance for Astroparticle Physics (HAP), the doctoral school KSETA at KIT, and Helmholtz Young Investigator Group (VH-NG-1055), Max Planck Research Group (MaxPlanck@TUM), and Deutsche Forschungsgemeinschaft DFG (Research Training Groups Grants No., GRK 1694 and GRK 2149, Graduate School Grant No. GSC 1085-KSETA, and SFB-1258 in Germany; Ministry of Education, Youth and Sport (CANAM-LM2015056, LTT19005) in the Czech Republic; Ministry of Science and Higher Education of the Russian Federation under contract 075-15-2020-778; and the Department of Energy through grants DE-FG02-97ER41020, DE-FG02-94ER40818, DE-SC0004036, DE-FG02-97ER41033, DE-FG02-97ER41041,  {DE-SC0011091 and DE-SC0019304 and the Federal Prime Agreement DE-AC02-05CH11231} in the United States. This project has received funding from the European Research Council (ERC) under the European Union Horizon 2020 research and innovation programme (grant agreement No. 852845).
We thank the computing cluster support at the Institute for Astroparticle Physics at Karlsruhe Institute of Technology, Max Planck Computing and Data Facility (MPCDF), and National Energy Research Scientific Computing Center (NERSC) at Lawrence Berkeley National Laboratory.




\newpage

\section*{Supplementary information}
\section{Combined analysis of first and second neutrino mass campaign}\label{sec:combined-analysis}
The 1000 days of total measurement time of the KATRIN experiment will be distributed within 15-20 individual measurement campaigns. The experimental parameters will not be constant over all these campaigns, as the systematic uncertainties are reduced and the experimental performance is improved (e.g. the background rate will be lowered) with time. Nevertheless, all individual campaigns need to be combined eventually to achieve the full statistical power.


{\subsection{Combination of individual results}
One approach is to simply multiply the \mnutwo\ distributions obtained with MC-propagation of the individual measurement phases. As the uncertainties of the first and second neutrino mass campaigns are still largely statistically dominated and the main systematic uncertainties are uncorrelated, this method is well justified for the combination presented here. The best fit, obtained from the combined distribution is $\mnutwo = \SI{0.11\pm0.34}{\electronvolt\squared}$. 
Assuming a symmetric Gaussian distribution of \mnutwo, we derive an upper limit of $\mnu < \SI{0.81}{\electronvolt}$ ( $\SI{90}{\percent}$ C.L). Adding the $\chi^2$-profiles of the individual fits leads to statistically consistent results. }


\subsection{Simultaneous fit of the both neutrino mass campaigns}
\begin{figure}
    \centering
    \includegraphics[width=\linewidth]{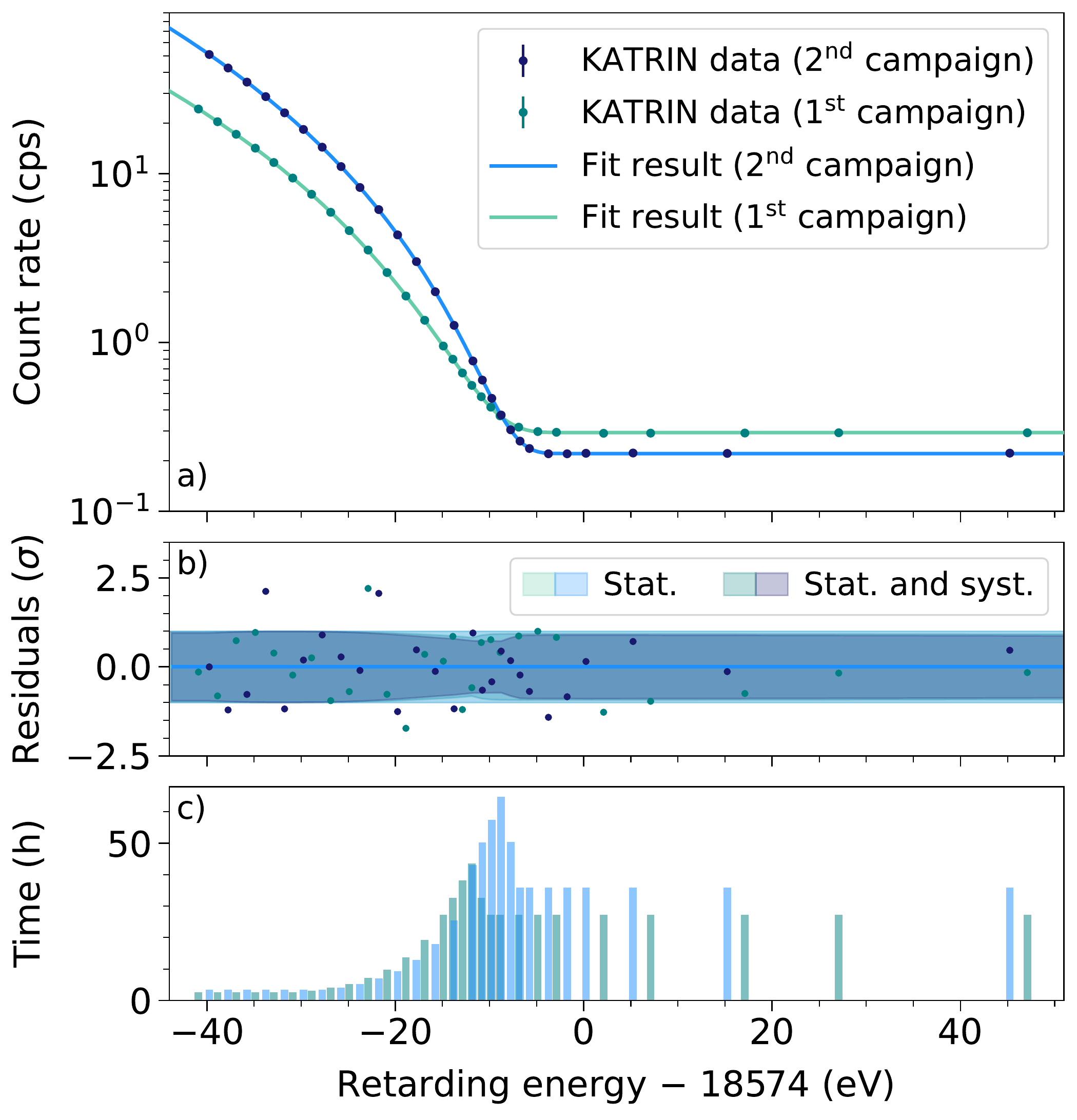}
    \caption{Simultaneous fit to the data of both the neutrino mass campaigns (KNM1 in green and KNM2 in blue) where the measured rates of the 12 detector rings are combined to a single detector area (uniform fit) in both campaigns. The combined best fit result is $\mnutwo = \SI{0.08 \pm 0.32}{\electronvolt\squared}$. b) Normalised residuals for the fit to the data. c) Measurement time distribution for KNM1 and KNM2. Both have 28 scan steps with slightly differing retarding energies. The most sensitive region of the spectrum where most of the measurement time is spent is shifted slightly to higher energies in the KNM2 campaign due to the improved signal to background.}
    \label{fig:combinedfit}
\end{figure}
Another way, to combine both data sets, is by performing a simultaneous fit of both data sets with a common neutrino mass. The first and second neutrino mass campaigns were performed under different experimental conditions, e.g. the column density was increased by a factor of four and the background level was reduced by \SI{25}{\percent} in the second campaign. Correspondingly, the systematic parameters and their uncertainties are partly different in the two campaigns. As a consequence, the data of the two campaigns cannot be described with a single effective spectrum prediction. Nevertheless, a simultaneous fit of both data sets can be performed by extending the $\chi^2$ function to
\begin{align}
\begin{split}
    \chi^2&(\mnutwo,\vec{\Theta}_{p1},\vec{\Theta}_{p2},\vec{\eta}_{p1},\vec{\eta}_{p2},\vec{\eta}_{c})   = \\  &\chi^2(\mnutwo,\vec{\Theta}_{p1},\vec{\eta}_{p1},\vec{\eta}_{c}) + \chi^2(\mnutwo,\vec{\Theta}_{p2},\vec{\eta}_{p2},\vec{\eta}_{c}) 
    + \mathrm{pull\, terms}
\end{split}
\label{eq:chi2}
\end{align}

where $\vec{\Theta}_{p1}$ and $\vec{\Theta}_{p2}$ depict the endpoint, background, and normalisation of the first ($p1$) and second ($p2$) measurement period. These parameters can be different in the two measurement phases, as the experimental conditions are changed. $\vec{\eta}_{p1}$ and $\vec{\eta}_{p2}$ correspond to the two sets of systematic parameters, while $\vec{\eta}_{c}$ illustrates the set of common systematic parameters. The neutrino mass squared \mnutwo\  is a common parameter of both $\chi^2$ functions. 

For a simultaneous fit the number of free parameters and data points increases by about a factor of two compared to the fit of a single measurement campaign,  which makes the minimisation computationally challenging and prone to numerical noise. To simplify the problem, we choose for this analysis to group the detector rings to one effective detector area (uniform fit), reducing the number of free parameters to $n_{\mathrm{free}}=33$ and the number of data points to $n_{\mathrm{data}}=56$.


With respect to the stand-alone first neutrino mass analysis as presented in \cite{aker2021analysis}, the spectrum calculation has been slightly improved. It now includes the transmission function for non-isotropic electrons, an improved parametrisation of the energy loss function, and a reduced uncertainty on the magnetic fields and column density. Finally, a possible Penning-trap induced time-dependent background, as described in the main text, is included. A re-analysis of the first neutrino mass data with these new inputs reveals a best fit of $\mnutwo = -1.0^{+0.9}_{-1.0}\,\mathrm{eV^2}$ and an upper limit of $\mnu < \SI{1.1}{\electronvolt}$ {at $\SI{90}{\percent}$ C.L}. These values are in good (\SI{10}{\percent}) agreement with respect to the published result in \cite{Aker:2019uuj}.

{The fit to both data sets is shown in Fig.\ \ref{fig:combinedfit}. The $\chi^2/\mathrm{ndof}= 51.7 / 49 =1.1$ indicates excellent agreement of the model to the data. The final result of the combined fit reveals $\mnutwo = \SI{0.08 \pm 0.32}{\electronvolt\squared}$ and a corresponding upper limit of $\mnu < \SI{0.76}{\electronvolt}$ {($\SI{90}{\percent}$ CL)}. }


In complement to the standard goodness-of-fit, we perform the Parameter-Goodness-of-Fit (PGoF) test~\cite{Maltoni:2003cu} to assess the compatibility of KNM1 and KNM2 data sets. This test quantifies the penalty of combining the data sets compared to fitting them independently. We find a PGoF probability of $\SI{17}{\percent}$ reflecting a good agreement between the two statistically independent data sets.


\subsection{Bayesian combination of both neutrino mass campaigns}

\begin{figure}[]
    \centering
    \includegraphics[width=\linewidth]{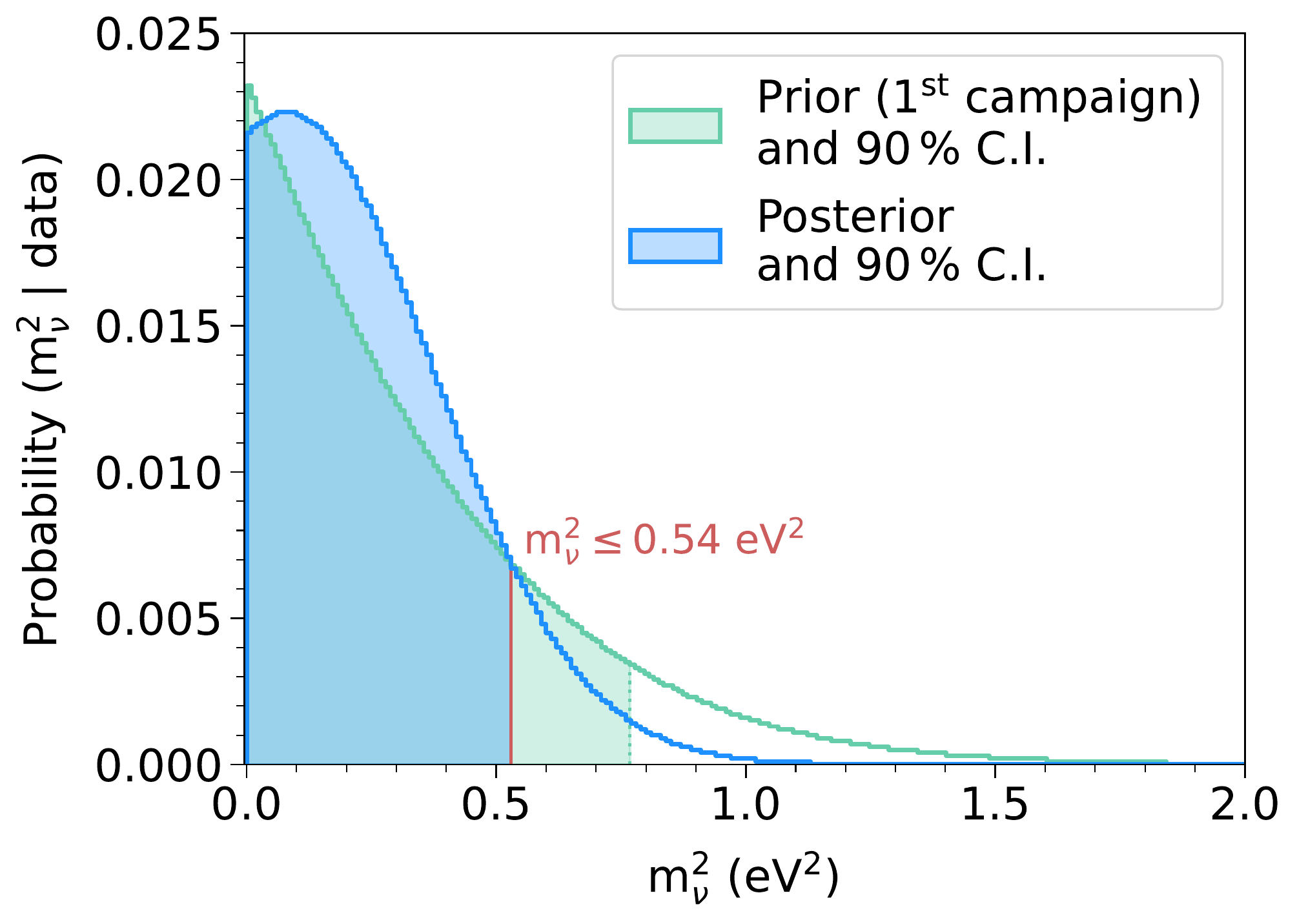}
   \caption{
   {Posterior distributions of the observable $\mnutwo$. The prior distribution for the KNM2 analysis (green)~\cite{aker2021analysis} is the posterior distribution from the KNM1 analysis. The resulting posterior distribution for the second campaign is shown in blue.}   }
    \label{fig:Bayesiancombi}
\end{figure}

Another way of combining different neutrino mass measurement campaigns is by using the posterior distribution of one campaign as prior information for the other campaign. In this procedure, correlations between the data sets are neglected.

For the analysis of the first neutrino mass measurement campaign, we use the result published in~\cite{Aker:2019nlx}, which includes the five major systematic effects. The resulting posterior distribution for \mnutwo, displayed in Fig.\  \ref{fig:Bayesian}, is then used as prior distribution of \mnutwo\ for the analysis of the second neutrino mass measurement campaign, which is otherwise performed with the same procedure as the stand-alone analysis.

The central value is found to be $\mnutwo=\SI{0.06\pm0.32}{\electronvolt\squared}$. The corresponding \SI{90}{\percent} credible interval from a positive prior on $\mnutwo$~corresponds to a Bayesian limit of $\mnu<\SI{0.72}{\electronvolt}$, as illustrated in Fig.\ \ref{fig:Bayesiancombi}.

\end{document}